\documentclass[reprint,amsmath,amssymb,aps,prxquantum]{revtex4-2}
 
\usepackage{graphicx}
\usepackage{dcolumn}
\usepackage{bm}
\usepackage{subcaption}
\usepackage{amsmath}
\usepackage{amssymb}
\usepackage{amsfonts}
\usepackage{color}
\usepackage{titlesec}
\newcounter{MEquation}

\titleformat{\section}[display]
{\large\bfseries}
{\thesection.}{0.5em}{}
\titlespacing{\section}{0em}{1em}{0em}

\titleformat{\subsection}[display]
{\normalfont\bfseries}
{\thesubsection.}{0.5em}{}
 \titlespacing{\subsection}{0em}{1em}{0em}
 
\titleformat{\subsubsection}[runin]
{\small\bfseries}
{\thesubsubsection.}{0.5em}{}
\titlespacing{\subsubsection}{0em}{0em}{1em}

\begin{document}

\title{Dynamical blockade of a reservoir for optimal performances of a quantum battery}

\author{F. Cavaliere$^{1,2}$, G. Gemme$^{1}$, G. Benenti$^{3,4}$, D. Ferraro$^{1,2,*}$,  and M. Sassetti$^{1,2}$}

\affiliation{$^{1}$ Dipartimento di Fisica, Università di Genova, Via Dodecaneso 33, 16146 Genova, Italy\\
$^{2}$ CNR-SPIN, Via Dodecaneso 33, 16146 Genova, Italy\\
$^{3}$ Center for Nonlinear and Complex Systems, Dipartimento di Scienza e Alta Tecnologia, Università degli Studi dell’Insubria, Via Valleggio 11, 22100 Como, Italy\\
$^{4}$ Istituto Nazionale di Fisica Nucleare, Sezione di Milano, Via Celoria 16, 20133 Milano, Italy\\
$^{*}$ e-mail: dario.ferraro@unige.it}

\date{\today}

\begin{abstract}
{The development of fast and efficient quantum batteries is crucial
for the prospects of quantum technologies. We show that both requirements are accomplished
in the paradigmatic model of a harmonic oscillator strongly coupled to a highly non-Markovian thermal reservoir. At short times, a {\em dynamical blockade} of the reservoir prevents the leakage of energy towards its degrees of freedom, promoting a significant accumulation of energy in the battery with high efficiency. The possibility of implementing these conditions in $LC$ quantum circuits opens up new avenues for solid-state quantum batteries.} 
\end{abstract}
\keywords{keywords}

\maketitle

The progressive broadening of the horizons of thermodynamics towards individual quantum mechanical systems out-of-equilibrium opened the way to the new field of quantum thermodynamics~\cite{Esposito09, Vinjanampathy16, Campisi16, Benenti17}. In this context, the possibility to properly characterize the {energetics} of miniaturized thermal machines represented a major boost for the development of quantum technologies devoted to energy storage and manipulation. Among them, quantum batteries (QBs) are currently assuming a pivotal role~\cite{Bhattacharjee21, Quach23}.

After the first appearance of the concept more than ten years ago~\cite{Alicki13}, these devices have been characterized by a frantic theoretical investigation~\cite{Campaioli23} followed recently by the first experimental realizations~\cite{Quach22}. The majority of the proposals discussed so far identify the QB as a collection of two-level systems charged by means of quantum~\cite{Alicki13, Binder15, Campaioli17, Ferraro18, Andolina18, Rossini20, Rosa20, Quach22, Crescente22, Mazzoncini23, Crescente24, Grazi24, Razzoli24b} or classical~\cite{Zhang19, Hu22, Gemme22, Gemme24} external sources. However, in recent years, the possibility to address multilevel QBs in view of achieving high storage capacity has been discussed~\cite{Seah21, Salvia23}.

Indubitably, among multilevel systems the quantum harmonic oscillator plays a prominent role due to its versatility. Indeed, a plethora of different physical systems can be ultimately described in terms of this simple and universal model. It is therefore not a surprise that also in the QB domain proposals for implementable multilevel devices based on quantum harmonic oscillators have appeared~\cite{Andolina18, Hovhannisyan20, Shaghaghi22, Shaghaghi23, Rodriguez23, Downing23, Rodriguez23b, Gangwar24}. Among the others, interesting schemes have considered the possibility to charge a harmonic oscillator QB 
by switching on the coupling with a reservoir for a suitable period 
of time~\cite{Farina19, Hovhannisyan20}. However, in previous investigations the efficiency of these protocols, defined as the ratio between the maximum extractable energy through unitary operations (known as ergotropy~\cite{Allahverdyan04}) and the energy required to switch on and off the coupling with the charger, remained very poor.

In this paper, 
{we show that this limitation can be overcome
and that both high efficiency and fast charging can be achieved, provided that: 
(i) the 
reservoir is engineered in a strongly non-Markovian regime characterized by a 
cut-off frequency $\omega_{\mathrm{D}}\gtrsim \omega_{0}$, with $\omega_{0}$ frequency of the oscillator and, counter-intuitively, (ii) the dissipation is very strong, with 
intensity $\gamma_{0}\gg \omega_{0}$.} In such a regime, at short enough times, the {switch-on coupling energy between QB and reservoir is almost completely transferred directly to the QB itself} with only marginal dissipation onto the reservoir degrees of freedom, due to a {\em dynamical blockade} of the reservoir. Consequently, the unit efficiency limit predicted for a QB and a reservoir initially in a thermal state~\cite{Barra22} can be saturated. 
{We explain this dynamical blockade in terms of} the hybridization between the QB and the reservoir modes with frequency close to a new emergent frequency $\Omega=\sqrt{\gamma_{0} \omega_{\mathrm{D}}}\gg \omega_{0}$. 
We point out that, in this underdamped regime, almost all the energy accumulated in the QB can be extracted. In addition, for sufficiently short times the energy oscillates back and forth in an almost periodic and coherent way  between the coupling energy and the QB, and the robustness of this effect over various periods of these energy oscillations can be exploited to mitigate the need for precise fine-tuning of the charging time. Furthermore, our results are astonishingly robust up to the thermal energies comparable with $\hbar \Omega$.

Due to the robustness of our protocol and the possibility to experimentally  
implement it by means of a quantum $LC$ circuit, playing the role of the QB,
embedded in a dissipative environment, with suitably engineered 
cut-off frequency and dissipation strength~\cite{Vool17, Blais21, Gramich11}, 
we strognly believe that the present study can open new and fascinating 
perspectives towards the realization of 
fast and highly efficient energy manipulation in solid-state devices.


\section*{Results}
\subsection*{Model of the quantum battery, charger and environment} 
\label{sec:model}
\noindent The QB is a quantum harmonic oscillator, strongly coupled to a many--modes reservoir described as a large collection of quantum harmonic oscillators in the framework of the Caldeira-Leggett model~\cite{Ingold02, Weiss_book,  Bhanja24}. The Hamiltonian of the system consists of three terms
\begin{equation}
\hat{H}^{(t)}=\hat{H}_{\mathrm{B}}+\hat{H}_{\mathrm{R}}+\theta(t)\hat{H}_{\mathrm{C}}\,,
\label{eq:H_tot}
\end{equation}
where $\theta(t)=0$ for times $t\leq 0$ while $\theta(t)=1$ for $t>0$. 
The QB is described by
\begin{equation}
\hat{H}_{\mathrm{B}}=\frac{\hat{p}^{2}}{2m}+\frac{m\omega^{2}_{0}}{2}\hat{x}^{2},
\label{eq:HB}
\end{equation}
with $\hat{x}$ and $\hat{p}$ position and momentum operators, and $m$ and $\omega_0$ respectively its mass and characteristic frequency. The reservoir Hamiltonian reads
\begin{equation}
\hat{H}_{\mathrm{R}}= \sum_{k} \hat{H}^{(k)}_{\mathrm{R}}=\sum_{k}\left(\frac{\hat{p}_{k}^{2}}{2 m_{k}}+\frac{m_{k} \omega^{2}_{k}}{2}\hat{x}_{k}^{2}\right)\,,
\label{eq:HR}
\end{equation}
where $\hat{x}_{k}$ and $\hat{p}_{k}$ are the position and momentum of the $k$--th mode, with mass $m_{k}$ and characteristic frequencies $\omega_{k}$. The QB and the reservoir are coupled through the term
\begin{equation}
\hat{H}_{\mathrm{C}}=\sum_{k} \hat{H}^{(k)}_{\mathrm{C}}=-\left(\sum_{k} c_{k}\hat{x}_{k}\right)\hat{x}+\frac{1}{2}m\Omega^2\hat{x}^2
\label{Eq:H_int}
\end{equation}
which, as will be shown, plays the role of a quantum charger. Eq.~(\ref{Eq:H_int}) is composed of two terms: a bilinear coupling between $\hat{x}$ and $\hat{x}_k$ -- with $c_k$ coupling constants which will be specified later on -- and a counter--term $\propto \hat{x}^2$, where
\begin{equation}
\label{eq:Omsq}
\Omega^2=\sum_{k}\frac{c^2_k}{mm_k\omega_k^2}\,,    
\end{equation}
which prevents the renormalization of the static potential of the QB~\cite{Ingold02, Weiss_book}.\\

The equations of motion for the QB and reservoir position operators in the Heisenberg picture read (from now on, $\hbar=k_{\mathrm{B}}=1$):
\begin{eqnarray}
\hat{\ddot{x}}(t)+(\omega_0^2+\Omega^2)\hat{x}(t)&=&\sum_{k}\frac{c_k}{m}\hat{x}_k(t)\,,\label{eq:EOMQB}\\
\hat{\ddot{x}}_k(t)+\omega_k^2\hat{x}_k(t)&=&\frac{c_k}{m_k}\hat{x}(t)\,.\label{eq:EOMQR}
\end{eqnarray}
Eq.~(\ref{eq:EOMQR}) can be formally solved considering $\hat{x}(t)$ a given external field, yielding
\begin{eqnarray}
\hat{x}_{k}(t)&=&\hat{x}_{k}(0)\cos\left(\omega_{k}t\right)+\frac{\hat{p}_k(0)}{m_{k}\omega_{k}}\sin\left(\omega_{k}t\right)\nonumber\\ 
&+&\frac{c_{k}}{m_{k} \omega_{k}}\int^{t}_{0} dt' \sin\left[\omega_{k}(t-t')\right]\hat{x}(t').
\label{eq:solxk}
\end{eqnarray}
Replacing this solution into Eq.~(\ref{eq:EOMQB}) one obtains the full operatorial quantum Langevin equation for the QB~\cite{Weiss_book,Carrega22,Cavaliere22}
\begin{equation}
\hat{\ddot{x}}(t)+\omega^{2}_{0}\hat{x}(t)+\int^{t}_{0} dt' \gamma(t-t')\hat{\dot{x}}(t')+\gamma(t) \hat{x}(0)=\frac{\hat{\xi}(t)}{m},
\label{eq:Lang}
\end{equation}
where the damping kernel  
\begin{equation}
\gamma(t)=\frac{1}{m}\theta(t)\sum_{k} \frac{c^{2}_{k}}{m_{k}\omega^{2}_{k}}\cos\left(\omega_{k}t\right)
\label{eq:gamma}
\end{equation}
and the reservoir fluctuating noise operator 
\begin{equation}
\hat{\xi}(t)=\sum_{k}c_{k} \left[ \hat{x}_{k}(0) \cos\left(\omega_{k}t\right)+\frac{\hat{p}_{k}(0)}{m_{k}\omega_{k}}\sin\left( \omega_{k}t\right)\right]
\label{eq:xi}
\end{equation}
have been introduced. Notice that the momentum operator is directly $\hat{p}(t)=m\hat{\dot x}(t)$.\\

The solution of Eq.~(\ref{eq:Lang}) can always be decomposed into a transient  homogeneous (h) and a non-homogeneous or thermal part (th):
\begin{equation}
\hat{x}(t)=\hat{x}_{\mathrm{h}}(t)+\hat{x}_{\mathrm{th}}(t).
\label{eq:decomposition}
\end{equation}
The first term is independent of the fluctuating field $\hat{\xi}(t)$ and is expressed via the response function $\chi(t)$ in terms of the initial conditions at time $t=0$ as
\begin{equation}
\hat{x}_{\mathrm{h}}(t)=\hat{\dot{x}}(0)\chi(t)+\hat{x}(0)\dot{\chi}(t)\,.
\label{eq:homogeneous}
\end{equation}
All the properties of $\chi(t)$, which satisfies initial conditions $\chi(0)=0$, $\dot{\chi}(0)=1$ and $\ddot{\chi}(0)=0$, are encoded into its Laplace transform (defined as $\hat{\tilde{x}}(\lambda)=\int^{+\infty}_{0} dt e^{-\lambda t} \hat{x}(t)$)
\begin{equation}
\tilde{\chi}(\lambda)=\frac{1}{\lambda^{2}+\omega^{2}_{0}+\lambda \tilde{\gamma}(\lambda)}\,,
\label{eq:chilaplace}
\end{equation}
where $\tilde{\gamma}(\lambda)$ is the Laplace transform of Eq.~(\ref{eq:gamma}).\\
The term $\hat{x}_{\mathrm{th}}(t)$ in Eq.~(\ref{eq:decomposition}) depends instead on the fluctuating field $\hat{\xi}(t)$ and is given by
\begin{equation}
\hat{x}_{\mathrm{th}}(t)=\int^{t}_{0} \frac{dt'}{m} \chi(t-t')\hat{\xi}(t')\,.
\label{non_homogeneous}
\end{equation}
\\

\subsection*{The charging/discharging protocol: energy balance and figures of merit}
\label{sec:energetics}
\noindent The charging/discharging protocol is comprised of four steps:\\

\noindent(\textbf{I}) At $t=0$, switching on $\hat{H}_{\mathrm{C}}$ the QB and the reservoir are connected;\\
\noindent(\textbf{II}) Energy flows for a {\em short} amount of time $t$;\\
\noindent(\textbf{III}) Switching off $\hat{H}_{\mathrm{C}}$, the QB and the reservoir are disconnected;\\
\noindent(\textbf{IV}) The maximum amount of energy is extracted from the QB, to be employed elsewhere.\\ 

A suitably engineered reservoir, especially within the protocol described above, will prove to be a valuable {\em resource} allowing fast and optimal charging of the QB. In the following  we will introduce the energy exchanges involved in the protocol and suitable figures of merit to quantify its performances.\\

\subsubsection*{\emph{Energy accumulated into the QB and ergotropy.}}
\label{sec:enerergo}
To begin with, let us introduce the energy {\em accumulated} into the QB over the charging time $t$ as~\cite{Ferraro18}
\begin{equation}
\Delta E_{\mathrm{B}}(t)=\langle \hat{H}_{\mathrm{B}}(t)\rangle-\langle \hat{H}_{\mathrm{B}}(0)\rangle 
\label{eq:E_B}
\end{equation}
where $\langle ... \rangle \equiv \mathrm{Tr}\left[...\hat{\rho}(0)\right]$ with $\hat{\rho}(0)$ the total initial density matrix of the system, which will be specified in the following, and where $\langle \hat{H}_{\mathrm{B}}(t)\rangle$ is the energy {\em stored} in the QB at time $t$. Here and in what follows, all operators evolve in the Heisenberg picture.\\

Clearly, not all the accumulated energy can actually be extracted from it. The {\em ergotropy} is defined as the maximum energy that can be extracted from the QB acting only with unitary operations~\cite{Allahverdyan04, Alicki13}. For a general setup the evaluation of the ergotropy is only possible via numerical approaches. However, for Gaussian states a closed expression is present~\cite{Farina19, Downing23}. To this effect we notice that the time evolution of the model under consideration maps Gaussian states at $t=0$ into Gaussian states at any later time $t$. Therefore, anticipating our choice of a Gaussian initial condition, the ergotropy of the QB at time $t$ is given by
\begin{equation}
\mathcal{E}(t)=\langle \hat{H}_{\mathrm{B}}(t)\rangle-\omega_{0}\sqrt{\det \sigma_{\mathrm{B}}(t)},
\label{eq:ergo}
\end{equation}
where $\sigma_{\mathrm{B}}(t)$ is the covariance matrix~\cite{Weedbrook12} at time $t$.\\

\subsubsection*{\emph{Work performed by the quantum charger.}}
Switching on and off of the quantum charger over a time $t$ requires a finite amount of work
\begin{equation}
W(t)=W_{\mathrm{on}}+W_{\mathrm{off}}(t)\,,
\label{eq:Wt}
\end{equation}
where
\begin{equation}
W_{\mathrm{on}}=\langle \hat{H}_{\mathrm{C}}(0)\rangle\quad;\quad W_{\mathrm{off}}(t)=-\langle \hat{H}_{\mathrm{C}}(t)\rangle\,,
\label{eq:Wonoff}
\end{equation}
with $\langle\hat{H}_{\mathrm{C}}(t)\rangle$ the coupling term in Eq.~(\ref{Eq:H_int}). 
Notice that, using the Langevin equation (\ref{eq:EOMQB}), $\langle\hat{H}_{\mathrm{C}}(t)\rangle$ can be expressed in terms of the QB variables only as
\begin{equation}
    \frac{\langle\hat{H}_{\mathrm{C}}(t)\rangle}{m}=\langle\hat{\dot{x}}^2(t)\rangle-\frac{1}{2}\frac{d^2}{dt^2}\langle\hat{x}^2(t)\rangle-\left(\omega_0^2+\frac{\Omega^2}{2}\right)\langle\hat{x}^2(t)\rangle\,.\label{eq:HCrewritten}
\end{equation}

In full analogy with what defined in Eq.~(\ref{eq:E_B}), the total work can be alternatively written as 
\begin{equation}
W(t)=-\left[\langle \hat{H}_{\mathrm{C}}(t)\rangle-\langle \hat{H}_{\mathrm{C}}(0)\rangle\right]=-\Delta E_{\mathrm{C}}(t)\,.
\end{equation}\\

\subsubsection*{\emph{Energy balance and spectral decomposition.}}
\label{sec:enbal}
Since the total system is closed, after the switching on, we can easily write the total energy balance  
\begin{equation}
\Delta E_{\mathrm{B}}(t)+ \Delta E_{\mathrm{R}}(t)= W(t)
\label{Balance}
\end{equation}
with a clear meaning: the total work $W(t)$ is distributed among the accumulated energy of the battery $\Delta E_{\mathrm{B}}(t)$ and the energy exchanged with the reservoir during the time $t$:
\begin{equation}
\Delta E_{\mathrm{R}}(t)=\langle\hat{H}_{\mathrm{R}}(t)\rangle-\langle\hat{H}_{\mathrm{R}}(0)\rangle\,.
\end{equation}
Eq.~(\ref{Balance}) further underlines the role of $\hat{H}_{\mathrm{C}}$ as a quantum charger. 
\begin{figure}[h!t]
    \includegraphics[width=0.45\textwidth]{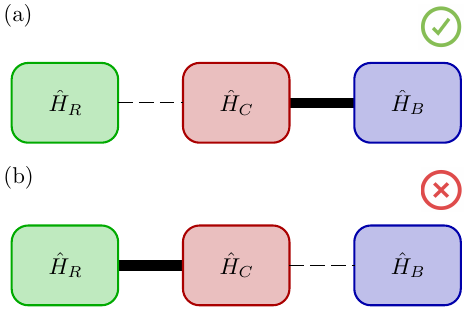}
    \caption{\textbf{Scheme of operation of $\hat{H}_{\mathrm{C}}$ as a quantum charger.} Panel (a) efficient and panel (b) inefficient charging protocol.}
    \label{fig:fig0}
\end{figure}
In this respect, an efficient charging protocol predominantly exchanges energy between the charger and the QB, see Fig.~\ref{fig:fig0}(a), while an inefficient protocol wastes most of its energy in the reservoir with little power delivered to the QB, as shown Fig.~\ref{fig:fig0}(b). Even though achieving an efficient charging seems almost hopeless in this context, given that the reservoir has a very large number of degrees of freedom which can collect energy with respect to the single one of the QB, we will show  that this is indeed possible via a careful engineering of the reservoir spectral density at strong coupling.\\

For later convenience we also introduce  the contributions to the charger and to the reservoir energy due to the $k$--th mode of the latter, defined as ($\nu=\mathrm{C,R}$)
\begin{equation}
\Delta E_{\nu}(t,\omega_{k})=\langle \hat{H}^{(k)}_{\nu}(t)\rangle-\langle \hat{H}^{(k)}_{\nu}(0^{})\rangle\label{eq:HCspec}
\end{equation}
which satisfy the sum rules
\begin{equation}
\Delta E_{\nu}(t)=\sum_{k} \Delta E_{\nu}(t,\omega_{k})\,.
\end{equation}
\\

\subsubsection*{\emph{Performances of the charging/discharging protocol.}}
In order to assess the performances of the charging procedure, two different  efficiencies are usually considered. The first is 
\begin{equation}
\eta_{\mathrm{B}}(t)=\frac{\mathcal{E}(t)}{\langle H_{\mathrm{B}}(t)\rangle}=1-\frac{\omega_{0}\sqrt{\det \sigma_{\mathrm{B}}(t)}}{\langle H_{\mathrm{B}}(t)\rangle}\,,
\label{eta_B}
\end{equation}
the ratio between the ergotropy and the energy stored in the QB. The second figure of merit, which is the most relevant when energy to connect and disconnect the charger is paid off ($W(t)>0$, see below and Ref.~\cite{Hovhannisyan20} for a discussion on this condition), is
\begin{equation}
\eta_{\mathrm{W}}(t)=\frac{\mathcal{E}(t)}{W(t)}\,
\label{eta_W}
\end{equation}
representing the ratio between the maximum work that can be extracted from the QB and the work paid to switch on and off the charger.

\subsection*{Engineered reservoir and operating regimes}
\label{sec:enbathopregs}
\subsubsection*{\emph{The response function.}}
\label{sec:resfun}
To proceed further we need to specify the couplings $c_k$ in the charging Hamiltonian $\hat{H}_{\mathrm{C}}$. Among the possible choices, we consider the paradigmatic case of Ohmic coupling which is the most common dissipation, especially when  the reservoir is represented by  quantum circuits \cite{Ingold92}. We have~\cite{Nieu02,Weiss_book,Grabert_physRe,Grabert_PRE} 

\begin{equation}
c_k=\omega_k\sqrt{\frac{2\gamma_0 m m_k\Delta}{\pi}\frac{\omega_{\mathrm{D}}^2}{\omega_{\mathrm{D}}^2+\omega_k^2}}\,.
\label{ckappa}
\end{equation}
Here, $\gamma_{0}$ is the coupling strength, $\omega_{\mathrm{D}}$ the Drude cut-off, and $\omega_k=k\Delta$ ($k$ positive integer) with $\Delta$ the constant level spacing of the modes of the reservoir. The behavior of the couplings defines the spectral density 
\begin{equation}
J(\omega)=\frac{\pi}{2} \sum_{k}\frac{c^{2}_{k}}{m_{k} \omega_{k}}\delta(\omega-\omega_{k})\,,
\label{J0}
\end{equation}
and the damping kernel -- see Eq.(\ref{eq:gamma}). Letting the number of modes to infinity, with $\Delta\to 0$, one eventually gets the continuum limit 
\begin{eqnarray}
&&J(\omega)=m \gamma_{0} \omega \frac{\omega^{2}_{\mathrm{D}}}{\omega^{2}+\omega^{2}_{\mathrm{D}}}
\label{J}\\
&&\gamma(t)=\gamma_{0} \omega_{\mathrm{D}}e^{-\omega_{\mathrm{D}}t}\theta(t)\ ;\quad\tilde{\gamma}(\lambda)=\frac{\gamma_0\omega_{\mathrm{D}}}{\lambda+\omega_{\mathrm{D}}}\,.
\label{eq:gamma_tL}
\end{eqnarray}
We can also identify the frequency scale $\Omega$ in Eq.~(\ref{eq:Omsq}) as
\begin{equation}
\Omega=\sqrt{\gamma_0\omega_{\mathrm{D}}}\,,
\label{eq:defOmega}
\end{equation}
which, as we will see, will play a crucial role on the charging dynamics of the QB.

\noindent As clear from the shape of $\gamma(t)$, the reservoir is highly non--Markovian when $\omega_{\mathrm{D}}$, the inverse of a memory time scale, is {\em small} $\omega_{\mathrm{D}}\approx\omega_0$, a regime of particular interest for this work. Such a regime has in the past drawn theoretical attention~\cite{Frigerio21,Wang01,Goychuk05,Hovhannisyan20}, and has recently been shown to be a resource in quantum thermodynamics~\cite{Razzoli24}. It is also important to stress that such a strong non--Markovian environment can nowadays be experimentally engineered in the context of quantum circuits~\cite{Cattaneo21,Gramich11,Vaaranta22,Kadijani20}.\\

We can now explicitly write the Laplace transform of the response function in Eq.~(\ref{eq:chilaplace})~\cite{Weiss_book}
\begin{equation}
\tilde{\chi}(\lambda)=\frac{\lambda+\omega_{\mathrm{D}}}{\left(\lambda^{2}+\omega^{2}_{0}\right)\left(\lambda+\omega_{\mathrm{D}}\right)+\lambda \gamma_{0}\omega_{\mathrm{D}}}
\label{eq:chi_lambda}
\end{equation}
which, in the time domain, can be written as
\begin{equation}
\chi(t)=\sum^{3}_{j=1}\chi_{j}e^{-\lambda_{j}t}\,.
\label{eq:chi}
\end{equation}
Here, $\lambda_j$ satisfies
\begin{equation}
\lambda^{3}-\lambda^{2}\omega_{\mathrm{D}}+\lambda\left(\omega^{2}_{0}+\gamma_{0}\omega_{\mathrm{D}}\right)-\omega_{\mathrm{D}}\omega^{2}_{0}=0,
\label{eq:polynomial}
\end{equation}
and
\begin{equation}
\chi_{j}=\left(\omega_{\mathrm{D}}-\lambda_{j}\right)\prod_{j'\neq j}\frac{1}{ \lambda_{j'}-\lambda_{j}}\,.
\label{eq:chi_j}
\end{equation}

\subsubsection*{\emph{Operating regimes.}}
\label{sec:opreg}
\begin{figure}[h]
    \includegraphics[width=0.45\textwidth]{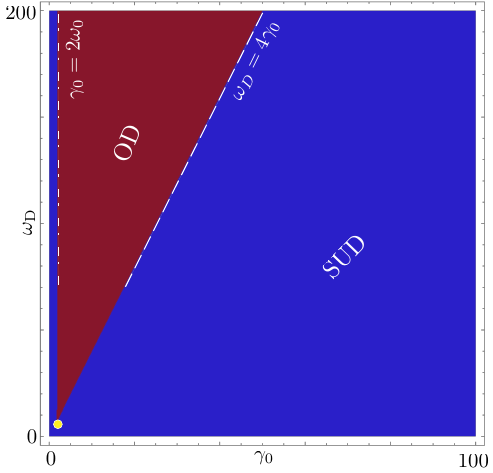}
    \caption{\textbf{Operating regimes.} As inferred by the nature of the zeroes of Eq.~(\ref{eq:polynomial}), as a function of $\gamma_{0}$ and $\omega_{\mathrm{D}}$ (units $\omega_{0}$), the red area denotes the overdamped (OD) regime, while the blue one corresponds to the underdamped (UD) one.  The bottom corner of the OD regime, marked as a yellow dot, has coordinates $\gamma_0=\frac{8}{3\sqrt{3}}\omega_0$ and $\omega_{\mathrm{D}}=3\sqrt{3}\omega_0$. The dash--dotted and dashed lines denote the boundaries between UD and OD  for $\omega_{\mathrm{D}}\gg\omega_0$.}
    \label{fig:fig1}
\end{figure}
As clear from Eq.~(\ref{eq:chi}), the nature of the roots $\lambda_j$ -- governed by the system parameters -- determines the characteristics of $\chi(t)$. From now on, we will elect $\omega_0$ as a typical energy scale and thus consider $\gamma_0$ and  $\omega_{\mathrm{D}}$ as free parameters. 
As shown in Fig.~\ref{fig:fig1}, due to the third degree polynomial with real coefficients, only two regimes can occur:\\

\noindent -- a \emph{underdamped} (UD) regime (blue part), characterized by one purely real and two complex conjugate roots.

\noindent -- a \emph{overdamped} (OD) regime (red part), where three real solutions occur. \\

To enter more into the details let us begin discussing the underdamped regime, by identifying the two representative regions. The first is the {\em weakly underdamped} one, with $\gamma_0\ll \omega_0,\omega_{\mathrm{D}}$, 
to the left of the OD part.
Here, to leading order in $\omega_{\mathrm{D}}$, $\chi(t)$ exhibits  oscillations at the bare frequency $\omega_0$, damped by the coupling with the reservoir~\cite{Weiss_book, Grabert_PRE}
\begin{equation}
\chi(t)\approx\frac{1}{\omega_0}e^{-\frac{\gamma_0}{2}t}\sin\left(\omega_0 t\right)\,.
\label{eq:chiUSUD0}
\end{equation}
This regime has already been discussed at large in the literature also about  QBs, typically employing a Lindblad master equation approach~\cite{Farina19, Rodriguez23, Qu23, Downing23}. Due to the small coupling, the  charging performances are here very poor. Therefore we are not going to investigate it in the rest of the paper.

\noindent The second region is to the right of OD at large coupling $\gamma_0\gg\omega_{\mathrm{D}}, \omega_0$: it corresponds to a  {\em strongly underdamped} (SUD) case. This region, has seldomly been discussed~\cite{Einsiedler20}, however it presents a peculiar behavior which, as we will see, will play a fundamental role in the QB dynamics. Here we have (see first part of the Methods for details)
\begin{equation}
\chi(t)\approx\frac{1}{\Omega}e^{-\frac{\omega_{\mathrm{D}}}{2}t}\sin\left(\Omega t\right)\,.
\label{eq:chiUSUD}
\end{equation}
As is clear, in this regime $\Omega=\sqrt{\gamma_0\omega_{\mathrm{D}}}$ takes the role of the new natural frequency of the QB. Notice that since the strong coupling  $\gamma_0\gg\omega_{\mathrm{D}},\omega_0$ implies $\Omega\gg\omega_{\mathrm{D}},\omega_0$,  this frequency  is strongly renormalized with respect to the bare one. In addition, its value is outside the band--width of the reservoir ($\Omega\gg\omega_{\mathrm{D}})$. As we will see, this point is crucial to achieve the best charging protocol, discussed in Fig.~\ref{fig:fig0} since the reservoir is dynamically blockaded around the frequency $\Omega$ of the QB preventing the energy absorption. 
For this reason in the rest of the paper we will mainly focus on this regime showing that it indeed produces the best short--time charging performances. 

\noindent We close commenting on the overdamped regime (red region), where $\chi(t)$ displays no oscillations. As a typical example we remind  the leading order behavior in $\omega_{\mathrm{D}}$, with $\gamma_0>2\omega_0$ (dashed-dotted line in the figure)~\cite{Weiss_book, Grabert_PRE}
\begin{equation}
\chi(t)\approx\frac{e^{-\frac{\gamma_0}{2}t}}{\sqrt{(\gamma_0/2)^2-\omega_0^2}}\sinh{\left[t\sqrt{(\gamma_0/2)^2-\omega_0^2}\right]}\,.
\label{eq:chiOD}
\end{equation}
Here, the dynamics of the QB is damped and, as shown in Supplementary Note 1, 
it is not very useful for our purposes since the reservoir absorbs the main part of the incoming energy with an inefficient charging protocol similar to Fig.~\ref{fig:fig0}(b). 

\subsection*{Main achievements}
\label{sec:results}
\subsubsection*{\emph{Relevant quantities.}}
\label{sec:incon}
To study in details the charging dynamics we start by specifying the initial conditions. We assume that, prior to the charge phase, the QB and the reservoir are disconnected. In addition, we consider an initial Gaussian state with the QB in its ground state $|g\rangle$ (i.e. a completely depleted battery) and the reservoir in its thermal equilibrium at temperature $T$. This corresponds to a factorized initial density matrix $\hat{\rho}(0)=\hat{\rho}_{\mathrm{B}}(0)\otimes \hat{\rho}_{\mathrm{R}}(0)$
with ($\beta=1/T$):
\begin{equation}
\hat{\rho}_{\mathrm{B}}(0)=|g\rangle \langle g|\ ;\ \hat{\rho}_{\mathrm{R}}(0)=\frac{e^{-\beta \hat{H}_{\mathrm{R}}(0)}}{\mathrm{Tr}\left\{e^{-\beta \hat{H}_{\mathrm{R}}(0)}\right\}}\,.
\label{eq:incon2}
\end{equation}
\noindent With this choice, the averages of position and momentum of both the QB and the reservoir variables are zero at $t=0$, while the second moments are $\langle\hat{x}^2(0)\rangle=\frac{1}{2m\omega_0}$, $\langle\hat{p}^2(0)\rangle=\frac{m\omega_0}{2}$, $\langle\{\hat{x}(0),\hat{p}(0)\}\rangle=0$, for the QB and  $\langle x_k(0)x_{k'}(0)\rangle=\frac{1}{2m\omega_k}\coth\left(\beta\omega_k/2\right)\delta_{k,k'}$, $\langle p_k(0)p_{k'}(0)\rangle=\frac{1}{2}m\omega_k\coth\left(\beta\omega_k/2\right)\delta_{k,k'}$, and $\langle\{x_k(0),p_{k'}(0)\}\rangle=0$ for the reservoir. Due to these  conditions the fluctuating noise $\xi(t)$ in Eq.(\ref{eq:xi}) has zero average $\langle\hat{\xi}(t)\rangle=0$ and time--translational invariant autocorrelation function~\cite{Weiss_book}
\begin{equation}
\!\!\!\!\!\langle\hat{\xi}(t)\hat{\xi}(0)\rangle\!\!=\!\!\int^{\infty}_{0}\!\!\frac{d \omega}{\pi} J(\omega)\!\!\sum_{n=\pm 1}\frac{e^{i n \omega t}}{2}\left[\coth\left(\frac{\beta\omega}{2}\right)-n\right]\!.
\label{eq:xi_correlator}
\end{equation}

\noindent Since the initial density is Gaussian, all previous considerations about the ergotropy hold true. All relevant quantities can be then evaluated from the covariance matrix $\sigma_{\mathrm{B}}(t)$, given here by
\begin{equation}
\sigma_{\mathrm{B}}(t)=m\begin{pmatrix}
\omega_0\langle\hat{x}^2(t)\rangle & \frac{1}{2}\frac{d}{dt}\langle\hat{x}^2(t)\rangle\\
&\\
\frac{1}{2}\frac{d}{dt}\langle\hat{x}^2(t)\rangle & \frac{1}{\omega_0}\langle\hat{\dot{x}}^2(t)\rangle
\end{pmatrix}.
\end{equation}
Much like the decomposition of $\hat{x}(t)$ into a homogeneous and thermal part, given in Eq.~(\ref{eq:decomposition}), the same holds for the covariance matrix: $\sigma_{\mathrm{B}}(t)=\sigma_{\mathrm{B,h}}(t)+\sigma_{\mathrm{B,th}}(t)$ where the $\mathrm{h}$ ($\mathrm{th}$) term only contains the homogeneous (thermal) contributions to $\langle\hat{x}^2(t)\rangle$ and $\langle\hat{\dot{x}}^2(t)\rangle$. 

\noindent The homogeneous terms are easily written in terms of the response function 
\begin{equation}    \begin{pmatrix}\langle\hat{x}_{\mathrm{h}}^2(t)\rangle\\\\\langle\hat{\dot{x}}_{\mathrm{h}}^2(t)\rangle\end{pmatrix}=\frac{\omega_0}{2m}\begin{pmatrix}\chi^2(t)\\\\\dot{\chi}^2(t)\end{pmatrix}+\frac{1}{2m\omega_0}\begin{pmatrix}\dot{\chi}^2(t)\\\\\ddot{\chi}^2(t)\end{pmatrix}\,.
\label{homo}
\end{equation}
More cumbersome are the thermal contributions, which can be written in the compact form 
\begin{eqnarray}    \!\!\!&&\begin{pmatrix}\langle\hat{x}_{\mathrm{th}}^2(t)\rangle\\\\\langle\hat{\dot{x}}_{\mathrm{th}}^2(t)\rangle\end{pmatrix}\!\!=\!\sum_{j,j'}\!\!\begin{pmatrix}1\\\\\lambda_j\lambda_{j'}\end{pmatrix}\!\!\int_{-\infty}^{\infty}\frac{d\omega}{2\pi m^2}\Biggl\{\frac{J(\omega)\coth\left(\frac{\beta\omega}{2}\right)}{(\lambda_j-i\omega)(\lambda_{j'}+i\omega)}\nonumber\\
&&\cdot\chi_j\chi_{j'}\left[1-e^{-(\lambda_j-i\omega)t}-e^{-(\lambda_{j'}+i\omega)t}+e^{-(\lambda_j+\lambda_{j'})t}\right]\Biggr\}\,.
\label{thermal}
\end{eqnarray}
Details are presented in the second part of the Methods {and in Supplementary Note 2}.\\

Concerning the efficiencies in  Eqs.~(\ref{eta_B}-\ref{eta_W}) we can deduce their  constraints. For $\eta_{\mathrm{B}}$ it follows directly that $\eta_{\mathrm{B}}(t)\leq 1$. Moreover, in the considered case, one also has $\eta_{\mathrm{W}}(t)\leq 1$ as can be seen reasoning as follows. Starting from Eqs.~(\ref{eq:ergo}) and (\ref{Balance}) one can write
\begin{eqnarray}
W(t)-\mathcal{E}(t)&=&\left[\omega_{0}\sqrt{\mathrm{det}\sigma_{\mathrm{B}}(t)}+\langle \hat{H}_{\mathrm{R}}(t) \rangle \right]\nonumber\\
&-&\left[\langle \hat{H}_{\mathrm{B}}(0) \rangle+\langle \hat{H}_{\mathrm{R}}(0) \rangle\right],
\end{eqnarray}
where the first line on the right hand side represents the total energy of QB+reservoir after the disconnection and the ergotropy extraction, while the second one is the initial internal energy. Assuming, consistently with what done in Eq.~(\ref{eq:incon2}), a passive initial state for the system and taking into account the fact that all the considered operations (time evolution of the system and ergotropy extraction) are unitary the internal energy can only increase or at most remain constant~\cite{Barra22}. Since $\mathcal{E}(t)\geq0$, one has $W(t)\geq \mathcal{E}(t)\geq 0$ finally proving the bound on $\eta_{\mathrm{W}}(t)$.\\

Finally, the (initial) work required to switch on the interaction (see Eq.~\ref{eq:Wonoff}) is 
\begin{equation}
W_{\mathrm{on}}=\frac{\gamma_0\omega_{\mathrm{D}}}{4\omega_0}=\frac{\Omega^2}{4\omega_0}\,.
\label{eq:maxen}
\end{equation}\\

\subsubsection*{\emph{The underdamped strong coupling regime.}}
We can now study the charging/discharging protocol, at {\em short times} in the underdamped strong coupling regime, which -- as will be shown -- provides the best performances. All results are obtained numerically. However, we will also provide analytical expressions to support our results. In addition, we will focus 
on custom tailored reservoirs with quite small $\omega_{\mathrm{D}}\gtrsim\omega_0$, i.e. with a narrow band and thus strongly non--Markovian, as this choice is one of the keys to obtain the best short--time performances among all the operating regimes.\\

\begin{figure}[h]
    \centering
    \includegraphics[width=0.45\textwidth]{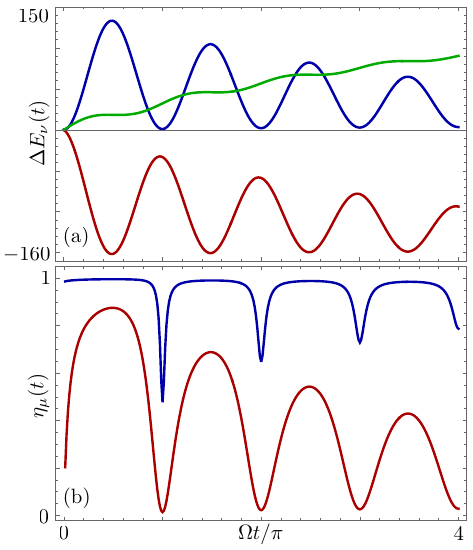}
    \caption{\textbf{Energetics and efficiency in the SUD regime.} Panel (a) plot of $\Delta E_{\nu}(t)$ (units $\omega_0$) with $\nu=\mathrm{B}$ (blue curve), $\nu=\mathrm{C}$ (red curve) and $\nu=\mathrm{R}$ (green curve) as a function of $\Omega t/\pi$; panel (b) plot of $\eta_{\mu}(t)$ with $\mu=\mathrm{B}$ (blue curve), and $\mu=\mathrm{W}$ (red curve) as a function of $\Omega t/\pi$. Parameters are $\omega_{\mathrm{D}}=2 \omega_{0}$, $\gamma_{0}=300 \omega_{0}$ (corresponding to $\Omega\approx24.5\omega_0$) and $T=0.1\omega_0$.}
    \label{fig:fig2}
\end{figure}

To begin our discussion, Fig.~\ref{fig:fig2}(a) shows the energy accumulated in the QB $\Delta E_{\mathrm{B}}(t)$ (blue line), the energy delivered by the charger $\Delta E_{\mathrm{C}}(t)=-W(t)$ (red line) and the energy dissipated into the reservoir $\Delta E_{\mathrm{R}}(t)$ (green line) as a function of time, in the quantum regime $T=0.1\omega_0$ for parameters within the SUD regime.\\

As is clear $\Delta E_{\mathrm{B}}(t)$ shows oscillations with a period $\pi/\Omega$, damped over a time scale $\omega_{\mathrm{D}}^{-1}$, and is locally maximum at times $t_{n}\approx\frac{\pi}{2\Omega}+n\frac{\pi}{\Omega}$, with $n\geq0$ an integer. Notice that this energy is large: in particular, the maximum amount accumulated in the QB, occurring at the shortest time $t=t_0\approx\pi/2\Omega$, is almost equal to $W_{\mathrm{on}}=\Omega^2/4\omega_0$. These are already hints of a fast and extremely effective charging protocol. Notice that the oscillations of $\Delta E_{\mathrm{B}}(t)$ are strikingly synchronized in phase opposition with respect to $\Delta E_{\mathrm{C}}(t)$, suggesting that at short times the charger and the QB almost perfectly exchange energy in lockstep. As a consequence, and as confirmed by the behaviour of $\Delta E_{\mathrm{R}}(t)$, during the first oscillations only a small fraction of energy is dissipated in the reservoir, whose dynamics seems to show an effective blockade at short times. We will turn back to this interpretation shortly.\\

The extreme effectiveness of the protocol is definitively confirmed by the behaviour of the efficiencies  $\eta_{\mathrm{B}}(t)$ and $\eta_{\mathrm{W}}(t)$, shown in panel (b). Except around the minima of $\Delta E_{\mathrm{B}}(t)$ we find $\eta_{\mathrm{B}}(t)\sim 1$ which implies the best ergotropy $\mathcal{E}(t)$: almost equal to the energy stored in the QB. This in turns implies that almost all the energy accumulated in the QB can be effectively retrieved to produce useful work. Most importantly, however, $\eta_{\mathrm{W}}(t)$ -- the ratio between the ergotropy and the switch on/off work -- is excellent: around the first maximum it is very close to the best attainable performance exceeding 0.85 (recall that in our case $\eta_{\mathrm{W}}(t)\leq 1$), and it is still very good around the fourth maximum where it reaches values above 0.5.\\ 

These results are indeed striking, especially in comparison with other similar charging-discharging protocols~\cite{Hovhannisyan20} that achieve significantly smaller values of $\eta_{\mathrm{W}}(t)$. In this respect, one common critique of a fast charging/discharging protocol -- such as the one shown here -- is the difficulty to fine--tune parameters in order to achieve the best performances~\cite{Hovhannisyan20}. However, in our scenario the almost perfect periodicity of all the quantities constitutes a key advantage: one is not forced to disconnect the QB from the charger at one peculiar time. Instead, the sequence of optimal times $t_n$ is highly predictable with charging energies and, as we have seen and as will be shown later on, performances remaining remarkable (and thus, stable) over several oscillation periods.\\

To close this part, we turn back to  interpret  this optimal charging protocol in terms of a dynamical blockade of the reservoir. We remind that the reservoir is heavily structured with a small band--width ($\omega_{\mathrm{D}}\ll\Omega$): this means that only the modes with $\omega_k\lesssim\omega_{\mathrm{D}}\ll \Omega$ have a significant weight in the spectral density $J(\omega)$ -- see Eqs.~(\ref{J0}) and (\ref{ckappa}). However, these modes are completely off resonance with the QB and then we expect that they can hardly absorb energy, thus allowing a back and forth energy transfer between QB and charger.
\begin{figure}[h]
    \centering
    \includegraphics[width=0.45\textwidth]{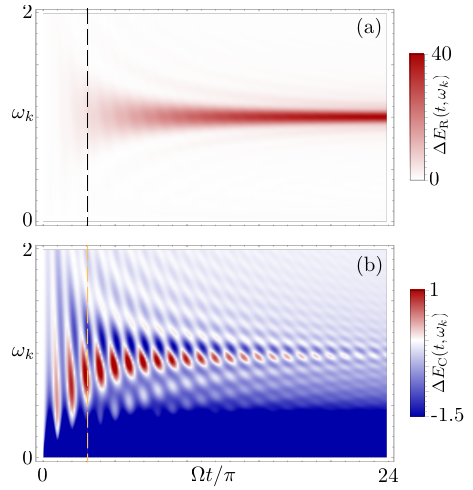}
    \caption{\textbf{Spectral decomposition of the reservoir and coupling energies in the SUD regime.} Density plots of (a) $\Delta E_{\mathrm{R}}(t,\omega_k)$ and (b) $\Delta E_{\mathrm{C}}(t,\omega_k)$, both in units $\Delta$, as a function of $\Omega t/\pi$ and $\omega_k$ (units $\Omega$). The dashed line represents the time window considered in previous Figures. Here $\omega_{\mathrm{D}}=2\omega_0$, $\gamma_0=300\omega_0$ ($\Omega\approx 24.5\omega_0$) and $T=0.1\omega_0$.}
    \label{fig:fig7}
\end{figure}

To demonstrate this scenario we inspect the spectral contributions of the reservoir $\Delta E_{\mathrm{R}}(t,\omega_k)$ defined in Eq.~(\ref{eq:HCspec}) -- see Supplementary Note 3 for details. Figure~\ref{fig:fig7}(a) shows a density plot of $\Delta E_{\mathrm{R}}(t,\omega_k)$, the dashed line represents the time window of Fig.~\ref{fig:fig2}. During the initial times the reservoir absorbs very little energy at all, due to the small value of the cut--off $\omega_{\mathrm{D}}\ll\Omega$.
This allows the important process of energy exchange between  the QB and the charger with anti--phase synchronization. This oscillating dynamics is clearly visible in the spectral contribution $\Delta E_{\mathrm{C}}(t,\omega_k)$ of the charger -- shown in panel (b) -- that oscillates between positive and negative values around $\omega_k=\Omega$. In particular the positive values signal that energy back--flows from the QB to the charger, since the energy in the reservoir slowly increases monotonically. On the other hand, away from the resonance $\Delta E_{\mathrm{C}}(t,\omega_k)<0$.

As time goes by, however, a sizeable amount of energy eventually accumulates into the reservoir: indeed the the latter starts to respond revealing a clear resonance precisely around the modes with $\omega_k\approx\Omega$. As already pointed out, modes at such frequencies have a very small weight in $J(\omega)$ but still succeed to open a narrow "energy pathway" into the reservoir -- albeit a slow one. In this time window the reservoir absorbs energy mainly at these frequencies, with all the other modes, including those at $\omega_k\lesssim\omega_{\mathrm{D}}$, contributing sensibly less. Ultimately, the fact that energy flows considerably into the reservoir only for times $t>\frac{\pi}{\Omega}$, and essentially only via this narrow channel, strongly slows down damping which in turns boosts the charging performances, promoting the back--and--forth exchange of energy between the charger and the QB at short times.\\

We will now study the stability of the above results  with respect to variations of $\Omega$ and $T$, showing that very good performances are achieved even if we choose less extreme parameters.

\begin{figure}[h!]
    \centering
    \includegraphics[width=0.45\textwidth]{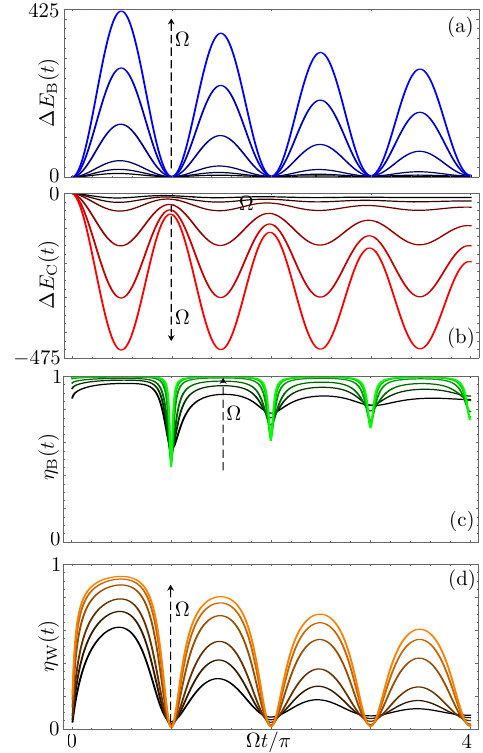}
    \caption{\textbf{Effects of $\Omega$.} Plot of (a) $\Delta E_{\mathrm{B}}(t)$ (units $\omega_0$), (b) $\Delta E_{\mathrm{C}}(t)$ (units $\omega_0$), (c) $\eta_{\mathrm{B}}(t)$, and (d) $\eta_{\mathrm{W}}(t)$ as a function of $\Omega t/\pi$ for $\omega_{\mathrm{D}}=2\omega_0$ and different values of $\gamma_0$: from black to color $\gamma_0=25\omega_0$ ($\Omega\approx7.1\omega_0$), $50\omega_0$ ($\Omega=10\omega_0$), $100\omega_0$ ($\Omega\approx14.1\omega_0$), $300\omega_0$ ($\Omega\approx24.5\omega_0$), $600\omega_0$ ($\Omega\approx 34.6\omega_0$) and $900\omega_0$ ($\Omega\approx42.4\omega_0)$. The dashed arrow marks the increasing direction of $\Omega$. In all panels $T=0.1\omega_0$.}
    \label{fig:fig3}
\end{figure}

We begin by varying $\Omega$, which we achieve by tuning $\gamma_0$ at fixed $\omega_{\mathrm{D}}$. Figure~\ref{fig:fig3} shows the results
for $\omega_{\mathrm{D}}=2\omega_0$ and low temperature $T=0.1\omega_0$. As can be seen, the qualitative behaviour is essentially unchanged within this wide range of $\Omega$, and the best charging performances always occur at shortest times. As shown in panels (a) and (c), the energy accumulated in the QB and $\eta_{\mathrm{B}}(t)$ {\em increase} for increasing $\Omega$. This implies that at strong(er) coupling more useful energy ($\eta_{\mathrm{B}}\to 1$) is stored in the QB. The almost out--of--phase behaviour of $\Delta E_{\mathrm{B}}(t)$ and $\Delta E_{\mathrm{C}}(t)$ is also confirmed -- see Panel (b). Also the efficiency $\eta_{\mathrm{W}}(t)$, shown in Panel (d), increases around the maxima for increasing coupling. It is even more important to observe, though, that even for the smallest value $\Omega\sim7\omega_0$ considered here, very good performances around the first maximum are achieved with a very good maximum for $\eta_{\mathrm{W}}(t)$ of about 0.6.

\begin{figure}[h!]
    \centering
    \includegraphics[width=0.45\textwidth]{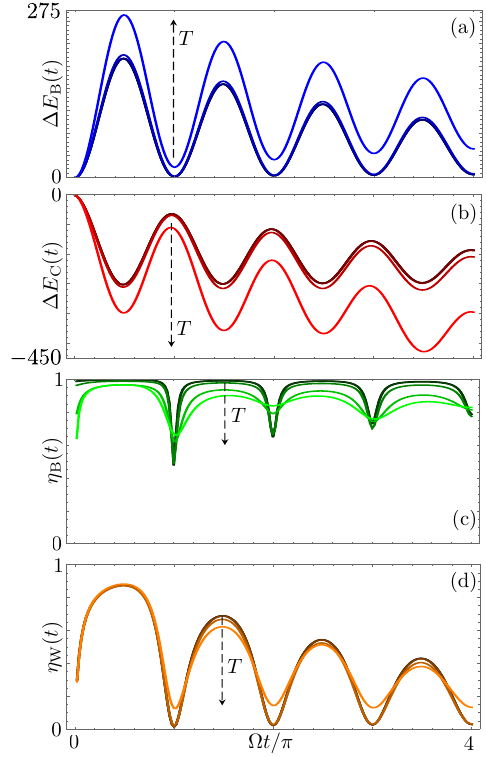}
    \caption{\textbf{Effects of $T$.} Plots of (a) $\Delta E_{\mathrm{B}}(t)$ (units $\omega_0$), (b) $\Delta E_{\mathrm{C}}(t)$ (units $\omega_0$), (c) $\eta_{\mathrm{B}}(t)$, and (d) $\eta_{\mathrm{W}}(t)$ as a function of $\Omega t/\pi$ for different values of $T$: from black to color $T=0.1\omega_0$ ($T\approx 4\cdot10^{-3}\Omega$), $\omega_0$ ($T\approx 4\cdot10^{-2}\Omega$), $10\omega_0$ ($T\approx 4\cdot10^{-1}\Omega$), $10^2\omega_0$ ($T\approx 4\Omega$), and $10^3\omega_0$ ($T\approx40\Omega$). The dashed arrow marks the increasing direction of $T$. In all panels $\omega_{\mathrm{D}}=2\omega_0$ and $\gamma_0=300\omega_0$.}    
    \label{fig:fig4}
\end{figure}


Concerning the effects of temperature, we show in Fig.~\ref{fig:fig4} the results for the same parameters of Fig.~\ref{fig:fig2} (SUD regime),  for temperatures ranging from very low to high ones. As can be seen, the behaviour of all the quantities of interest is essentially independent of the temperature and only when $T>\Omega$ deviations are found. In particular, notice that even though at the largest temperatures considered here $\eta_{\mathrm{B}}(t)$ slightly decreases, its value around the maxima still is $\gtrsim 0.8$, which means that a high ergotropy can be retrieved from the QB even in this regime. Also, when $T>\Omega$ the oscillations of $\eta_{\mathrm{W}}(t)$ tend to become slightly less pronounced as $T$ increases, although we stress that the first maximum exhibits an exceptional resilience.\\

These results show a strong stability  with respect to $\Omega$ and $T$. This leads to the important conclusion that our fast charge protocol is within reach of state of the art experimental platforms, without requiring exceedingly fast control or exceedingly low temperatures, see discussion about experimental feasibility below.\\

In the SUD regime, when $\Omega\gg\omega_{\mathrm{D}},\omega_0$ and in the short times limit $t\lesssim n\pi/\Omega$ (with $n>1$ an integer of order unity), analytic expressions for $\langle\hat{H}_{\mathrm{B}}(t)\rangle$, the ergotropy and for the total work $W(t)$ can be obtained, that support our findings. Deferring all derivations to Supplementary Note 2, here we just quote the main results valid {\em near the local maxima} of $\Delta E_{\mathrm{B}}(t)$.\\ 



\noindent At low temperature all quantities are dominated by homogeneous terms and one finds, to leading order in $\Omega$, that
\begin{equation}
\langle\hat{H}_{\mathrm{B}}(t)\rangle\approx\langle\hat{H}_{\mathrm{B}}(t)\rangle_{\mathrm{h}}\ ;\ \mathcal{E}(t)\approx\langle\hat{H}_{\mathrm{B}}(t)\rangle_{\mathrm{h}}\ ;\ W(t)\approx W_{\mathrm{h}}(t)\,,\nonumber
\end{equation}
with
\begin{equation}
\label{eq:EnAsLow}
\langle\hat{H}_{\mathrm{B}}(t)\rangle_{\mathrm{h}}=\frac{\Omega^2}{4\omega_0}e^{-\omega_{\mathrm{D}} t}\sin^2(\Omega t)\,.
\end{equation}
Notice that in this regime the ergotropy is optimal, since it essentially matches the energy stored in the QB. One also has
\begin{equation}
W_{\mathrm{h}}(t)=\frac{\Omega^2}{4\omega_0}\left[1-e^{-\omega_{\mathrm{D}} t}\cos^2(\Omega t)\right]\,,
\label{eq:WlowT}
\end{equation}
which allows to obtain a closed expression for the efficiency
\begin{equation}
\eta_{\mathrm{W}}(t)=\frac{e^{-\omega_{\mathrm{D}} t}\sin^2(\Omega t)}{1-e^{-\omega_{\mathrm{D}} t}\cos^2(\Omega t)}\,.
\end{equation}
These equations provide results in excellent agreement with the numerical ones (see Supplementary Note 2).\\


\noindent Conversely at high temperature ($T\gg\omega_{\mathrm{D}}$) the energy of the QB, the ergotropy and the work acquire  thermal contributions in addition to the homogeneous ones in Eqs.~(\ref{eq:EnAsLow})  and (\ref{eq:WlowT}). We find  $\langle\hat{H}_{\mathrm{B}}(t)\rangle=\langle\hat{H}_{\mathrm{B}}(t)\rangle_{\mathrm{h}}+\langle\hat{H}_{\mathrm{B}}(t)\rangle_{\mathrm{th}}$, with
\begin{equation}
\langle\hat{H}_{\mathrm{B}}(t)\rangle_{\mathrm{th}}=\frac{T}{2}\Big[e^{-\omega_{\mathrm{D}} t}\sin^2(\Omega t)+(1-e^{-\omega_{\mathrm{D}} t})\Big]\,,
\end{equation}

Concerning the total work $W(t)=W_{\mathrm{h}}(t)+ W_{\mathrm{th}}(t)$ we have
\begin{equation}
\!\!\!\!\!W_{\mathrm{th}}(t)=\frac{T}{2}\left[e^{-\omega_{\mathrm{D}} t}\sin^2(\Omega t)+\frac{\Omega^2}{\omega_0^2}\left(\!1-e^{-\frac{2\omega_0^2\omega_{\mathrm{D}} t}{\Omega^2}}\right)\right].
\end{equation}
 Comparing the additional thermal parts with respect to the homogeneous ones we see how the former become effectively dominant only for $T>\Omega$. The situation is slightly more complicated for the ergotropy $\mathcal{E}(t)$, since at high temperatures its expression deviates from that of $\langle\hat{H}_{\mathrm{B}}(t)\rangle_{\mathrm{h}}$ -- see Fig.~\ref{fig:fig4}(c) -- signalling that the contribution of the covariance matrix becomes non negligible -- see Eq.~(\ref{eq:ergo}). One can still obtain an analytical expression for this quantity by evaluating the corrections due to the latter term starting from the approximated form of the variances reported in Eqs.~(S1) and~(S16) of the Supplementary Note 2. The final form is however too cumbersome to be reported.\\
\noindent All the above expressions are in excellent agreement with our numerical results (see Supplementary Note 2).\\

\noindent To close this Section, it is interesting to answer the question: what would happen, in the SUD regime, if one would defer the disconnection of the charger to very long times $t\to+\infty$?  Notice that in this regime the reduced density matrix of the QB is given by the trace over the reservoir~\cite{Weiss_book} $\hat{\rho}_{\mathrm{B}}(+\infty)=\mathrm{Tr}_{\mathrm{R}}\left\{\frac{e^{-\beta\hat{H}}}{Z}\right\}$, where $Z=\mathrm{Tr}\left\{e^{-\beta\hat{H}}\right\}$.
Leaving all analytical details to Supplementary Note 2, here we quote the main results for both low and the high temperatures.\\

At low temperatures one finds $\langle\hat{H}_{\mathrm{B}}(+\infty)\rangle\approx\mathcal{E}\approx\frac{\Omega}{4}$ and $W(+\infty)\approx\frac{\Omega^2}{4\omega_0}$. Charging is still possible, and the ergotropy remains comparable with the energy stored in the QB. Observe however that in this case the latter quantity is $\propto\Omega$, while in the quick charge protocol outlined above the QB energy and ergotropy are of the order of $\Omega^2$, see Eq.~(\ref{eq:EnAsLow}). Thus, in this case much less energy can be effectively stored and retrieved. Even more importantly, in this regime charging requires a very large work so that one has $\eta_{\mathrm{W}}(+\infty)\approx\omega_0/\Omega\ll1$ and thus a much poorer efficiency with respect to the short--time performances.\\

In the high temperature regime ($T\gg\Omega$), dominated by energy equipartition, the situation is even worse. Indeed one finds $\langle\hat{H}_{\mathrm{B}}(+\infty)\rangle\approx T$ but, crucially, $\mathcal{E}\propto T^{-1}$ -- implying $\eta_{\mathrm{W}}(+\infty)\to 0$. This means that no useful work can be extracted from the QB: the charging is useless.\\

Even more, also for other parameters the long--time performances never come even close to those obtained at short times in the SUD regime.\\

\subsubsection*{\emph{Considerations about experimental feasibility.}}
\label{sec:expfeas}
We conclude this part commenting about the fact that the ideal platform to test the functioning of the discussed device is represented by a quantum $LC$ circuit, playing the role of the QB, embedded in an dissipative environment. In particular, considering for example values of $L\approx 500\,\, \mathrm{nH}$ for the inductance and $C\approx 500 \,\,\mathrm{pF}$ for the capacity this circuit can be described as a quantum harmonic oscillator with characteristic frequency $\omega_{0}/2\pi\approx 10\,\,\mathrm{MHz}$ (with $\omega_{0}=1/\sqrt{LC}$)~\cite{Vool17, Blais21}. Focusing on this frequency regime, which is typically smaller with respect to what usually considered in circuit QED experiments~\cite{Naik17}, has the practical advantage to allow a control of the device on a time scale in the nanosecond range which is compatible with the discussed features of the figures of merits (the first maximum of $\eta_{\mathrm{W}}$ occurring at $t_{0}\approx 1\,\,\mathrm{ns}$ assuming for example $\gamma_{0}=300 \omega_{0}$ and $\omega_{\mathrm{D}}=2 \omega_{0}$ as in Fig.~\ref{fig:fig2}). The possibility to observe the discussed phenomenology is also assured by its great stability with respect to thermal effect as long as $k_{\mathrm{B}}T\lesssim \hbar \Omega$ (see Fig.~\ref{fig:fig4}). Within the discussed range of parameters this leads to the threshold temperature $T\approx 12\,\,\mathrm{mK}$ compatible with the cryogenic temperatures typically reached in quantum transport and quantum computing experiments carried out in solid-state devices.\\

Concerning the properties of the reservoir, a simple lumped parameters model comprises a resistive part $R_E$ and a capacitive part $C_E$~\cite{Gramich11}. At small frequencies $<\omega_{\mathrm{D}}$, the resistive part dominates and one is left with a simple series RLC circuit. From the equation of motion for the charge $Q$ of such a circuit
\begin{equation}
\ddot{Q}+\frac{R_E}{L} \dot{Q}+\frac{1}{LC}Q=0\,, 
\label{eq:rlc}
\end{equation}
comparing to the homogeneous part of Eq.~(\ref{eq:Lang}) it is possible to identify the charge $Q$ with $\hat{x}$ as frequently done in the framework of the quantum circuits~\cite{Ingold02}. Then, Eq.~(\ref{eq:rlc}) allows to deduce that 
\begin{equation}
R_E\approx \gamma_{0} L.
\end{equation}
With the parameters considered above, 
this leads to $R\approx 10^{4}\, \Omega$ which is of the same order of magnitude of the resistance quantum usually appearing in solid-state devices. The simple model outlined above breaks down at higher frequencies $\gtrsim\omega_{\mathrm{D}}$, when the capacitive effects of the environment become relevant. They induce a typical cut-off frequency~\cite{Gramich11} of the order of $(R_E C_E)^{-1}$, which we identify with $\omega_{\mathrm{D}}$. The parameters considered in this text allow to estimate $C_E\approx 1\,\,\mathrm{pF}$. To conclude we note that quantum couplers can be used to realize a fast control of the coupling between the QB and the charger~\cite{Sete21, Campbell23, Heunisch23}.

\section*{Conclusions}
\label{sec:conclusions}
\noindent In this paper we have proposed a {scheme for} a quantum battery based on a quantum harmonic oscillator, strongly coupled to a non--Markovian reservoir via a quantum charger. The {considered} procedure relies on the transient dynamics which occurs right after the battery is connected to the quantum charger, thus ensuring a {\em quick charging} protocol. We have shown that the evolution of the energy stored in the battery is almost periodic, which allows to avoid a too precise fine-tuning of the time at which the battery {need to} be disconnected from the charge. Moreover, we have shown that this protocol is {\em very efficient}, allowing 
in principle to extract through unitary operations practically all the energy stored in the quantum battery, with a ratio between the energy that can be extracted and the work done by the charger which approaches the ideal unit limit.
These outstanding features are due to two key ingredients, namely the {\em non-Markovianity} due to a reservoir with a spectral density with a cut-off of the order of the oscillator frequency, and a peculiar -- and as yet almost unexplored -- {working regime in the} {\em underdamped regime at strong coupling}. Crucially, these result in a {\em dynamical blockade} of the reservoir dynamics, which allows an almost coherent exchange of energy between the charger and the quantum battery {at short enough times}.\\
Such a protocol may be envisioned with a quantum $LC$ circuit playing the role of the battery and with the required environment being suitably engineered via state--of--the--art quantum circuits, and thus may genuinely contribute to a significant advancement in the field of quantum energy storage {in solid state devices}.\\

This work will pave the way for new developments in the field of quantum energy routing and management, opening for instance the possibility to study the charging of, and energy transfer between, two or more quantum batteries strongly coupled to a highly non--Markovian reservoir acting as a energy bus or energy router.


\section*{Methods}

\small
\subsection*{The response function $\chi(t)$ in the strong underdamped regime}
\label{App:app2}
\noindent In this part we give some details on the behavior of the response function $\chi(t)$.\\

\noindent The cubic equation in Eq.~(\ref{eq:polynomial})
\begin{equation}
\lambda^{3}-\lambda^{2}\omega_{\mathrm{D}}+\lambda\left(\omega^{2}_{0}+\gamma_{0}\omega_{\mathrm{D}}\right)-\omega_{\mathrm{D}}\omega^{2}_{0}=0
\label{appeq:polynomial}
\end{equation}
yields the general Vieta's relations

\begin{eqnarray}
\lambda_{1}+\lambda_{2}+\lambda_{3}&=&\omega_{\mathrm{D}}\,,\label{VietaUno}\\
\lambda_{1}\lambda_{2}+\lambda_{1}\lambda_{3}+\lambda_{2}\lambda_{3}&=&\omega^{2}_{0}+\gamma_{0}\omega_{\mathrm{D}}\,,\\
\lambda_{1}\lambda_{2}\lambda_{3}&=&\omega_{\mathrm{D}}\omega^{2}_{0}\,.\label{VietaTre}
\end{eqnarray}
We remind that, for finite damping, all roots have positive real parts. In addition, in the underdamped regime two roots are complex and one is real, so they can be written as 
\begin{equation}
    \lambda_{1,2}=\Gamma\pm i\nu, \quad 
     \lambda_3\in\mathbb{R}\,,
\end{equation}
where $\Gamma,\nu\in\mathbb{R}$.
Using Eqs.~(\ref{VietaUno}) and~(\ref{VietaTre}) we have
\begin{equation}
    \Gamma=\frac{\omega_{\mathrm{D}}-\lambda_3}{2}, \quad
    \Gamma^2+\nu^2=\frac{\omega_{\mathrm{D}}\omega_0^2}{\lambda_3}.
    \label{app:lambda12}
\end{equation}
According to this, the $\chi_{j}$ defined in Eq.~(\ref{eq:chi_j}) are
\begin{eqnarray}
    \chi_1&=&\frac{i}{2\nu}\left[\frac{2 \nu +i \left(\lambda_3+\omega_{\mathrm{D}}\right)}{2 \nu +i \left(3\lambda_3-\omega_{\mathrm{D}}\right)}\right], \quad \chi_2=\chi_1^*\\
    \chi_3&=&4\frac{\omega_{\mathrm{D}}-\lambda_{3}}{\left(\omega_{\mathrm{D}}-3\lambda_3\right)^2+4\nu^2}.
\end{eqnarray}
We now derive analytic expressions at strong coupling with $\gamma_0\gg\omega_{\mathrm{D}}, \omega_0$, that is $\Omega=\sqrt{\gamma_0\omega_{\mathrm{D}}}\gg \omega_0,\omega_{\mathrm{D}}$.
Here, to leading order in $\Omega$, the real the solution of Eq.~(\ref{appeq:polynomial}) is 
\begin{equation}
    \lambda_3\approx\frac{\omega_0^2\omega_{\mathrm{D}}}{\Omega^2}.
    \label{app:lambda3}
\end{equation}
Inserting this results in Eqs.~(\ref{app:lambda12}) one has
\begin{equation}
    \Gamma\approx\frac{\omega_{\mathrm{D}}}{2}, \quad
    \nu\approx \Omega\, .
    \label{app:lambda12bis}
\end{equation}
Concerning $\chi_{j}$  we find 
\begin{equation}
    \!\!\!\chi_1\approx\frac{i}{2 \Omega}\left[1+i\frac{\omega_{\mathrm{D}}}{\Omega}\right],\,\,
    \chi_2=\chi_1^*,\,\,
    \chi_3\approx\frac{\omega_{\mathrm{D}}}{\Omega^2}.
    \label{app:chij}
\end{equation}
This leads to the following expression for the response function $\chi(t)=\sum_{j}\chi_j e^{-\lambda_j t}$
\begin{equation}
\chi(t)\approx \frac{e^{-\frac{\omega_{\mathrm{D}}}{2} t}}{\Omega}
\left[\sin{\left(\Omega t\right)}-\frac{\omega_{\mathrm{D}}}{\Omega}\cos{\left(\Omega t\right)}\right]+\frac{\omega_{\mathrm{D}}}{\Omega^2}e^{-\frac{\omega_0^2\omega_{\mathrm{D}}}{\Omega^2} t}\,,
\label{app:chi_big_gamma0}
\end{equation}
whose leading order expansion is 
\begin{equation}
\chi(t)\approx \frac{e^{-\frac{\omega_{\mathrm{D}}}{2} t}}{\Omega}
\sin{\left(\Omega t\right)}
\label{app:chi_big_gamma}
\end{equation}
as quoted in Eq.~(\ref{eq:chiUSUD}).

\subsection*{Explicit expression for QB variances}\label{App:app3}
In this part we give details on how to get the general expressions of the variances $\langle x^2(t)\rangle$ and $\langle \dot{x}^2(t)\rangle$ quoted in the main text -- see Eqs.~(\ref{homo}) and~(\ref{thermal}).

\noindent We start by recalling that the solution of the Langevin equation~(\ref{eq:Lang}) can be decomposed as $\hat{x}(t)=\hat{x}_{\mathrm{h}}(t)+\hat{x}_{\mathrm{th}}(t)$: this will be the building block for evaluating the above quantities. The homogeneous part is given by 
\begin{equation}
\hat{x}_{\mathrm{h}}(t)=\hat{\dot{x}}(0)\chi(t)+\hat{x}(0)\dot{\chi}(t)\,,
\label{app:homogeneous}
\end{equation}
while the thermal one is 
\begin{equation}
\hat{x}_{\mathrm{th}}(t)=\int^{t}_{0} \frac{dt'}{m} \chi(t-t')\hat{\xi}(t')\, ,
\label{app:non_homogeneous}
\end{equation}
which, due to the initial condition $\chi(0)=0$ implies also 
\begin{equation}
\hat{\dot x}_{\mathrm{th}}(t)=\int^{t}_{0} \frac{dt'}{m} \dot\chi(t-t')\hat{\xi}(t')\,.
\label{app:non_homogeneousdot}
\end{equation}
Here, the dot represents the derivative with respect to the first argument.

\noindent On a very general ground, given initial decoupled conditions and $\langle \hat{\xi}(t)\rangle=\langle \hat{x}(t)\rangle=0$,\,$\langle \hat{\dot x}(t)\rangle=0$,  the homogeneous and thermal terms factorize as 
\begin{eqnarray}
\langle \hat{x}^2(t)\rangle&=&\langle \hat{x}^2_{\mathrm{h}}(t)\rangle+\langle \hat{x}^2_{\mathrm{th}}(t)\rangle\,, \nonumber\\
\langle \hat{\dot x}^2(t)\rangle&=&\langle \hat{\dot x}^2_{\mathrm{h}}(t)\rangle+\langle \hat{\dot x}^2_{\mathrm{th}}(t)\rangle\,.
\label{app:factorization}
\end{eqnarray}
These matrix elements completely define the contributions to the covariance matrix
$\sigma_{\mathrm{B}}(t)=\sigma_{\mathrm{B,h}}(t)+\sigma_{\mathrm{B,th}}(t)$ given by
\begin{equation}
\sigma_{\mathrm{B,h/th}}(t)=m\begin{pmatrix}
\omega_0\langle\hat{x}_{\mathrm{h/th}}^2(t)\rangle & \frac{1}{2}\frac{d}{dt}\langle\hat{x}_{\mathrm{h/th}}^2(t)\rangle\\
&\\
\frac{1}{2}\frac{d}{dt}\langle\hat{x}_{\mathrm{h/th}}^2(t)\rangle & \frac{1}{\omega_0}\langle\hat{\dot{x}}_{\mathrm{h/th}}^2(t)\rangle
\end{pmatrix}\,.
\label{app:sigmah}
\end{equation}
We now observe that the homogeneous part is independent of the temperature and can be easily written in term of the response function  $\chi(t)$. Taking into account the initial conditions and  using  Eq.~(\ref{app:homogeneous})
one straightforwardly arrives to the expressions 
\begin{equation}    \begin{pmatrix}\langle\hat{x}_{\mathrm{h}}^2(t)\rangle\\\\\langle\hat{\dot{x}}_{\mathrm{h}}^2(t)\rangle\end{pmatrix}=\frac{\omega_0}{2m}\begin{pmatrix}\chi^2(t)\\\\\dot{\chi}^2(t)\end{pmatrix}+\frac{1}{2m\omega_0}\begin{pmatrix}\dot{\chi}^2(t)\\\\\ddot{\chi}^2(t)\end{pmatrix}\,.
\label{app:homo}
\end{equation}
As expected, $\sigma_{\mathrm{B,h}}(t\to\infty)=0$.\\

Concerning the thermal contributions we need a more careful manipulation of Eqs.~(\ref{app:non_homogeneous}) and (\ref{app:non_homogeneousdot}). In the following we describe how to explicitly write $\langle \hat{x}_{\mathrm{th}}^{2}(t)\rangle$.

\noindent Using Eq.~(\ref{app:non_homogeneous})
we have
\begin{equation}
\langle \hat{x}_{\mathrm{th}}^{2}(t)\rangle=\frac{1}{m^{2}}\int^{t}_{0}dt_{1} \int^{t}_{0}dt_{2} \chi(t-t_{1}) \chi(t-t_{2})\langle\hat{\xi}(t_{1})\hat{\xi}(t_{2}) \rangle 
\label{app:xthintegral}
\end{equation}
Since the integrand is symmetric under exchange of $t_2$ with $t_1$, only the symmetric part of the correlation function $\langle\hat{\xi}(t_{1})\hat{\xi}(t_{2})\rangle$ contributes. 
We call it~\cite{Cavaliere23,Carrega24}
\begin{equation}
\mathcal{L}_s(t_1-t_2)=\!\!\int^{\infty}_{0}\!\!\frac{d \omega}{\pi} J(\omega)\coth\left(\frac{\beta\omega}{2}\right)\cos\left[\omega(t_1-t_2)\right]\,,
\label{app:xi_correlator}
\end{equation}
see Eq.~(\ref{eq:xi_correlator}), with Fourier transform 
\begin{equation}
\mathcal{L}_s(\omega)=J(\omega)
\coth\left(\frac{\beta \omega}{2}\right).
\end{equation}
Inserting now the general expression of $\chi(t)$, given in Eq.~(\ref{eq:chi}), and integrating over time one finally obtains 
\begin{eqnarray}
&&\langle \hat{x}_{\mathrm{th}}^{2}(t)\rangle 
= \sum^{3}_{j,j'=1} \chi_{j}\chi_{j'}  \int^{+\infty}_{-\infty} \frac{ d \omega}{2 \pi m^{2}} \frac{\mathcal{L}_s(\omega)}{(\lambda_{j}-i\omega)(\lambda_{j'}+i \omega)} \nonumber\\&&\left[1-e^{-\left(\lambda_{j}-i \omega\right)t} -e^{-\left(\lambda_{j'}+i \omega\right)t}+e^{-\left( \lambda_{j}+\lambda_{j'}\right)t}\right].
\label{app:xquadro}
\end{eqnarray}

\noindent The last term to be evaluated is 
 $\langle \hat{\dot x}^2_{\mathrm{th}}(t)\rangle$.  This follows straightforwardly by inserting the expression in Eq.~(\ref{app:non_homogeneousdot}) which gives
 \begin{equation}
\langle \hat{\dot x}_{\mathrm{th}}^{2}(t)\rangle=\frac{1}{m^{2}}\int^{t}_{0}dt_{1} \int^{t}_{0}dt_{2} \dot \chi(t-t_{1}) \dot\chi(t-t_{2})\langle\hat{\xi}(t_{1})\hat{\xi}(t_{2}) \rangle\,. 
\label{app:xdotthintegral}
\end{equation}
Following similar steps as done for $\langle \hat{x}_{\mathrm{th}}^{2}(t)\rangle$ a form similar to Eq.~(\ref{app:xquadro}) is found, except for the substitutions $\chi_j\cdot\chi_{j'}\to\lambda_{j}\chi_j\cdot\lambda_{j'}\chi_{j'}$ -- see also Eq.~(\ref{thermal})
\begin{eqnarray}
&&\langle \hat{\dot x}_{\mathrm{th}}^{2}(t)\rangle 
= \sum^{3}_{j,j'=1} \lambda_{j}\chi_{j}\lambda_{j'}\chi_{j'}  \int^{+\infty}_{-\infty} \frac{ d \omega}{2 \pi m^{2}} \frac{\mathcal{L}_s(\omega)}{(\lambda_{j}-i\omega)(\lambda_{j'}+i \omega)} \nonumber\\&&\left[1-e^{-\left(\lambda_{j}-i \omega\right)t} -e^{-\left(\lambda_{j'}+i \omega\right)t}+e^{-\left( \lambda_{j}+\lambda_{j'}\right)t}\right].
\label{app:xdotquadro}
\end{eqnarray}

We close this part by commenting on the long time limit. For $t\rightarrow+\infty$
only the first term in the squared brackets of $\langle \hat{x}_{\mathrm{th}}^{2}(t)\rangle$ and $\langle \hat{\dot x}_{\mathrm{th}}^{2}(t)\rangle$ survives (see Eqs.~(\ref{app:xquadro}) and (\ref{app:xdotquadro})) since ${\rm Re}\{\lambda_j\}>0$. Then, after some manipulations, we arrive to the form 
\begin{eqnarray}
\langle \hat{x}_{\mathrm{th}}^{2}(+\infty)\rangle&=&\int^{+\infty}_{-\infty}\frac{ d \omega}{2 \pi m}\coth\left(\frac{\beta \omega}{2}\right) {\rm Im}\left[\tilde{\chi}\left(\lambda=-i \omega\right)\right]\, ,
\label{x_th_infinite}\nonumber\\
\langle \hat{\dot{x}}_{\mathrm{th}}^{2}(+\infty)\rangle &=&\!\int^{+\infty}_{-\infty}\!\!\!\frac{ d \omega }{2 \pi m}\omega^{2}\coth\left(\frac{\beta \omega}{2}\right) {\rm Im}\left[\tilde{\chi}\left(\lambda=-i \omega\right) \right]\,.\nonumber\\
\end{eqnarray}
In addition, we can easily see from Eq.~(\ref{app:xquadro})
that 
\begin{equation}
\lim_{t\to\infty}\frac{d}{dt}\hat{x}_{\mathrm{th}}^{2}(t)= 0\,.
\label{app:xdotx}
\end{equation}
This confirms the expected fluctuation-dissipation theorem~\cite{Weiss_book}. 

\small
\section*{Data availability}
All data is available in the manuscript and in the Supplementary Information.

\bibliographystyle{naturemag}
\bibliography{biblio}

\small
\section*{Acknowledgments}
\noindent D.F. acknowledges the contribution
of the European Union-NextGenerationEU through the
``Quantum Busses for Coherent Energy Transfer'' (QUBERT) project, in the framework of the Curiosity Driven 2021 initiative of the University of Genova. G.B, F.C. and D.F. acknowledge support from the project PRIN 2022 - 2022XK5CPX (PE3) SoS-QuBa - "Solid State Quantum Batteries: Characterization and Optimization" and M.S. acknowledges support from the project PRIN 2022 - 2022PH852L (PE3) SoS-QuBa - "Non reciprocal supercurrent and Topological Transition in hybrid Nb-InSb nanoflags", both projects  funded within the programme "PNRR Missione 4 - Componente 2 - Investimento 1.1 Fondo per il Programma Nazionale di Ricerca e Progetti di Rilevante Interesse Nazionale (PRIN)", funded by the European Union - Next Generation EU.".
\small
\section*{Author Contributions}
\noindent D.F., M.S. and  G.B. conceived the idea. D.F., G.G. and M.S. developed the analytical calculations. F.C. developed the numerical calculations and wrote the first version of the manuscript. G.B., D.F. and G.G. revised the manuscript and gave final approval for publication.
 
\section*{Competing interests}
\noindent The authors declare no competing interests.

\section*{Additional Information}
\noindent Supplementary information.
\end{document}


\title{Supplementary Information:\\Dynamical blockade of a reservoir for optimal performances of a quantum battery}

\author{F. Cavaliere$^{1,2}$, G. Gemme$^{1}$, G. Benenti$^{3,4}$, D. Ferraro$^{1,2,*}$,  and M. Sassetti$^{1,2}$}

\affiliation{$^{1}$ Dipartimento di Fisica, Università di Genova, Via Dodecaneso 33, 16146 Genova, Italy\\
$^{2}$ CNR-SPIN, Via Dodecaneso 33, 16146 Genova, Italy\\
$^{3}$ Center for Nonlinear and Complex Systems, Dipartimento di Scienza e Alta Tecnologia, Università degli Studi dell’Insubria, Via Valleggio 11, 22100 Como, Italy\\
$^{4}$ Istituto Nazionale di Fisica Nucleare, Sezione di Milano, Via Celoria 16, 20133 Milano, Italy\\
$^{*}$ e-mail: dario.ferraro@unige.it}
\maketitle

\section*{Supplementary Note 1: The overdamped regime}
\label{App:appOD}
This part is devoted to  study the QB performance in the overdamped regime.  The most relevant physical difference with respect to the SUD regime, discussed in the main text, is that here the reservoir has a band--width $\omega_{\mathrm{D}}$ larger than the damping $\gamma_0$,  i.e. there are many more modes of the environment sizably coupled to the charger in comparison with the previous regime.
\begin{figure}[h!]
    \centering
    \includegraphics[width=0.45\textwidth]{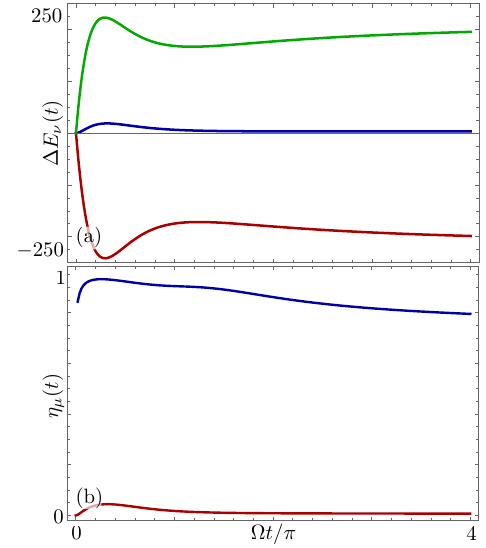}
    \caption{\textbf{Energetics and efficiency in the OD regime.} Plots of (a) $\Delta E_{\nu}(t)$ with $\nu=\mathrm{B}$ (blue curve), $\nu=\mathrm{C}$ (red curve) and $\nu=\mathrm{R}$ (green curve) as a function of $\Omega t/\pi$; (b) $\eta_{\mu}(t)$ with $\mu=\mathrm{B}$ (blue curve), and $\mu=\mathrm{W}$ (red curve) as a function of $\Omega t/\pi$. In panel (a) $\Delta E_{\nu}(t)$ is given in units $\omega_0$, parameters for all panels are $\omega_{\mathrm{D}}=60 \omega_{0}$, $\gamma_{0}=10 \omega_{0}$ and $T=0.1\omega_0$.}
    \label{fig:fig5}
\end{figure}
This in turns leads to a marked deterioration of the charging performances. To make a fair comparison, in Fig.~\ref{fig:fig5} we show the same quantities as in Fig.~3 in the main text
with the same value of   $\Omega\approx24.5\omega_0$ - equal initial work $W_{\mathrm{on}}$ -  but for $\omega_{\mathrm{D}}=60\,\omega_0$ and $\gamma_0=10\,\omega_0$, well within the OD regime.\\

The differences are striking. Inspecting panel (a) no quantity oscillates as expected, and $\Delta E_{\mathrm{B}}(t)$ reaches a maximum of only $\sim W_{\rm on}/6$, a worse performance with respect to the SUD case. Even more strikingly, $\Delta E_{\mathrm{C}}(t)$ and $\Delta E_{\mathrm{R}}(t)$ proceed in almost perfect anti-phase, implying a massive leakage of energy from the charger to the reservoir -- see Fig.~1(b) in the main text.

\noindent This has a direct impact on the charging performance. The ratio $\eta_{\mathrm{B}}(t)$ remains large, although worse than in the SUD regime, but in striking comparison to the latter case now $\eta_{\mathrm{W}}(t)<0.1$ as shown in panel (b), in accordance with the fact that most of the work done by the charger is dissipated into the reservoir. Instead, in contrast to the SUD regime, it is now the dynamics of the QB to be essentially blocked. As a consequence, for fixed $\gamma_0$ the QB essentially absorbs the same amount of energy regardless how large is $\omega_{\mathrm{D}}$ and this implies that more and more energy is wasted into the reservoir.\\

\begin{figure}[h]
    \centering
    \includegraphics[width=0.45\textwidth]{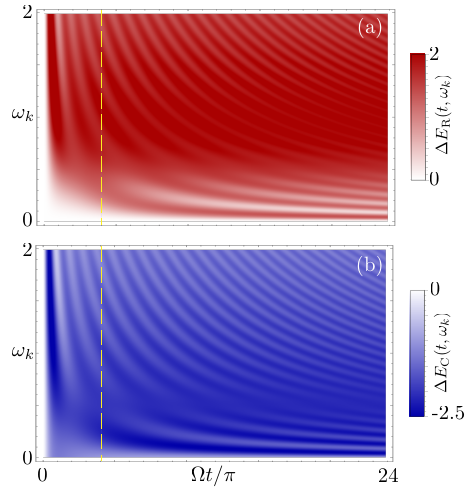}
    \caption{\textbf{Spectral decomposition of the reservoir and coupling energies in the OD regime.} Density plots of (a) $\Delta E_{\mathrm{R}}(t,\omega_k)$ and (b) $\Delta E_{\mathrm{C}}(t,\omega_k)$, both in units $\Delta$, as a function of $\Omega t/\pi$ and $\omega_k$ (units $\Omega$). The dashed line represents the time window considered in previous Figures. Parameters are the same as in Fig.~\ref{fig:fig5}.}
    \label{fig:fig8}
\end{figure}

To support these conclusions, Fig.~\ref{fig:fig8} shows the spectral decompositions $\Delta E_{\mathrm{R}}(t,\omega_k)$ and $\Delta E_{\mathrm{C}}(t,\omega_k)$. Panel (a) confirms that no resonance is found in the modes of the reservoir. Instead, the latter receives energy over a very wide spectrum of energies, up to $\omega_{\mathrm{D}}\gg\Omega$: a positive inflow of energy to the reservoir is found which is matched, as shown in panel (b), by the global outflow of energy from the charger -- the majority of which is directly dissipated into the reservoir. 

\begin{figure}[h]
    \centering
    \includegraphics[width=0.45\textwidth]{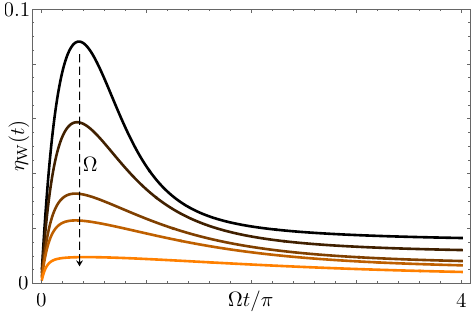}
    \caption{\textbf{Effects of $\Omega$ in the OD regime.} Plot of $\eta_{\mathrm{W}}(t)$ as a function of $\Omega t/\pi$ for $\gamma_0=6\omega_0$ and different values of $\omega_{\mathrm{D}}$: from black to color $\omega_{\mathrm{D}}=30\omega_0$, $50\omega_0$, $100\omega_0$, $150\omega_0$, $200\omega_0$. The dashed arrow marks the increasing direction of $\Omega$. Here $T=0.1\omega_0$.}
    \label{fig:fig6}
\end{figure}

Figure~\ref{fig:fig6} further supports the fact that in the OD regime there is a blockade of the QB dynamics. Here, $\eta_{\mathrm{W}}(t)$ is shown for fixed $\gamma_0=6\omega_0$ and different values of $\omega_{\mathrm{D}}$. As $\Omega$ is increased the charging efficiency gets worse and worse, in accordance with the fact that more and more energy is routed towards the reservoir.\\

\section*{Supplementary Note 2: Analytic results at strong underdamped coupling regime}
\label{App:app4}
In this part we discuss the analytic results obtained in the strong underdamped regime ($\Omega=\sqrt{\gamma_0\omega_{\mathrm{D}}}\gg \omega_0,\omega_{\mathrm{D}}$), both at short and long times.\\

\centerline{\bf a. Short times}

We start by considering the contributions to $\langle x^2(t)\rangle$ and $\langle \dot{x}^2(t)\rangle$ of the homogeneous part, quoted in Eq.~(73) of the main text.
As one can see, everything depends on the response function $\chi(t)$, which we will now discuss in the strong underdamped regime. Inserting Eq.~(67) of the main text and taking the leading order contributions in $\Omega$,  we arrive to these dominant terms for each component
\begin{SMEquations}    \begin{pmatrix}\langle\hat{x}_{\mathrm{h}}^2(t)\rangle\\\\\langle\hat{\dot{x}}_{\mathrm{h}}^2(t)\rangle\end{pmatrix}\!\!\approx\frac{1}{2m\omega_0}\begin{pmatrix}\dot{\chi}^2(t)\\\\\ddot{\chi}^2(t)\end{pmatrix}\approx\frac{e^{-\omega_{\mathrm{D}} t}}{2m\omega_0}\begin{pmatrix}\cos^2{(\Omega t)}\\\\\\\Omega^2\sin^2{(\Omega t)}\end{pmatrix}\,.
\label{app:homo1}
\end{SMEquations}

\noindent As clear, both  variances oscillate with frequency $\Omega$ and are weakly damped ($\omega_{\mathrm{D}}\ll \Omega$). For large $\Omega$, $\langle\hat{\dot{x}}^2(t)\rangle$ dominates over $\langle{{x}}^2(t)\rangle$ (at least away from its minima).

\noindent As already said, these terms are independent on temperature and decay to zero at long times $t\to\infty$. However, as shown in the main text, they play a dominant role at short times .
To discuss this point in details we need to estimate also the thermal contributions. 

\noindent We start by considering Eqs.~(74)  
and~(78) of the main text 
at short times: namely $\Omega t\lesssim n\pi/2$ (with $n$ integer of order unity), which represent the time interval of the first QB oscillations.  
We now rewrite the symmetric kernel $\mathcal{L}_s(t)$ in a compact form, to extract the scaling in $\Omega$. Using the Drude spectral function -- see Eqs.~(75) 
 and~(30) of the main text
\begin{SMEquations}
\mathcal{L}_s(t)=\frac{m\omega_{\mathrm{D}}\Omega^2}{\pi}F\left(\omega_{\mathrm{D}} t\right)
\label{app:Lcompact}
\end{SMEquations} 
where 
\begin{SMEquations}
F(s)=\int^{+\infty}_{0}\frac{x\ dx}{1+x^2}\coth\left(\frac{\omega_{\mathrm{D}}}{2T}x \right)\cos(xs)\,.
\label{app:F}
\end{SMEquations} 
 We can now write the variances  $\langle\hat{x}_{\mathrm{th}}^2(t)\rangle$ and $\langle\hat{\dot x}_{\mathrm{th}}^2(t)\rangle$
by rescaling the variables according to $x_{i}=\Omega t_{i}$ ($i=1,2$) and introducing Eq.~(\ref{app:Lcompact}) for $\mathcal{L}_s(t)$.  We have 

\begin{SMEquations}  \begin{pmatrix}\langle\hat{x}_{\mathrm{th}}^2(t)\rangle\\\\\langle\hat{\dot{x}}_{\mathrm{th}}^2(t)\rangle\end{pmatrix}=
\frac{\omega_{\mathrm{D}}}{\pi m}
\int^{\Omega t}_{0}\!\!\!\!dx_{1} \int^{\Omega t}_{0}\!\! \!\!dx_{2}F\left[\frac{\omega_{\mathrm{D}}}{\Omega}(x_2-x_1)\right]\begin{pmatrix}\chi\left(\frac{x_{1}}{\Omega}\right)\chi\left(\frac{x_{2}}{\Omega}\right) 
\\\\\Omega^2\frac{d}{dx_1}\chi\left(\frac{x_{1}}{\Omega}\right)\frac{d}{dx_2}\chi\left(\frac{x_{2}}{\Omega}\right) \end{pmatrix}\,.
\label{app:integralwithF}
\end{SMEquations}
Notice that $x_{1,2}$ are at most of order unity, since we consider short times $\Omega t\approx O(1)$,  
and then $\omega_{\mathrm{D}}(x_2-x_1)/\Omega\ll 1$. We now discuss the two opposite limits of low and high temperatures.\\

\noindent {\bf \em -- Low temperatures ($T\ll\omega_{\mathrm{D}}$)}\\

\noindent In this regime $\coth\left(\frac{\omega_{\mathrm{D}}}{2T}x \right)\to 1$ and we have
\begin{SMEquations}
F(s)\approx -\frac{1}{2}\sum_{p=\pm 1}e^{ps}{\rm Ei}(-ps)\,,
\end{SMEquations}
which in the regime $\omega_{\mathrm{D}} t\ll 1$ (namely, $s\ll 1$), considered here, becomes
\begin{SMEquations}
F(s)\approx -\big[{\cal C}+\ln (|s|) \big]\,,
\label{app:FlowT}
\end{SMEquations}
\!\!with $\cal C$ the Euler constant. Now, in order to estimate how the integrals (\ref{app:integralwithF}) scales with 
$\Omega$  we exploit the dominant behavior of $\chi(t)\approx e^{-\frac{\omega_{\mathrm{D}}}{2}t}\sin(\Omega t)/\Omega$. 
As can be seen, the leading  scaling in $\Omega$ is given by 
\begin{SMEquations}  \begin{pmatrix}\langle\hat{x}_{\mathrm{th}}^2(t)\rangle\\\\\langle\hat{\dot{x}}_{\mathrm{th}}^2(t)\rangle\end{pmatrix}\propto\frac{\omega_{\mathrm{D}}}{m\,\Omega^2}
\ln\left({\frac{\Omega}{\omega_{\mathrm{D}}}}\right)
\begin{pmatrix}1
\\\\\Omega^2\end{pmatrix}\,,
\end{SMEquations} 
\!\!\!\! where the proportionality factor is at best of order one. 
This behavior is definitely less dominant than 
the one obtained for the homogeneous part given by (see Eqs.(\ref{app:homo1}))
\begin{SMEquations}    \begin{pmatrix}\langle\hat{x}_{\mathrm{h}}^2(t)\rangle\\\\\langle\hat{\dot{x}}_{\mathrm{h}}^2(t)\rangle\end{pmatrix}\!\!\propto\frac{1}{m\omega_0}\begin{pmatrix}1\\\\\\\Omega^2
\end{pmatrix}\,.\qquad
\end{SMEquations}
\!\!\!\! These results demonstrate that, at low temperatures, the homogeneous part alone well describes the behavior of the two variances and then of the quantities of interest.  

\noindent We close this discussion by summarizing the expected behaviors of the different figures of merits in the strong coupling regime and at low temperatures. As of the energy stored in the QB the first maxima are given by 
\begin{SMEquations}
\langle\hat{H}_{\mathrm{B}}(t)\rangle\approx\langle\hat{H}_{\mathrm{B}}(t)\rangle_{\mathrm{h}}=\frac{\Omega^2}{4\omega_0}e^{-\omega_{\mathrm{D}} t}\sin^2(\Omega t)\,.
\label{app:hb}
\end{SMEquations}
\!\!\!\! Concerning the ergotropy $\mathcal{E}(t)=\langle\hat{H}_{\mathrm{B}}(t)\rangle - 
\omega_{0}\sqrt{\det\sigma_{\mathrm{B}}(t)}$,   we 
notice that $\sigma_{\mathrm{B}}(t)=\sigma_{\mathrm{B,h}}(t)+\sigma_{\mathrm{B,th}}(t)$, defined in Eq.~(72) of the main text, is never dominant at large $\Omega$. Then, around the maxima of the QB, we have 
\begin{SMEquations}
\mathcal{E}(t)\approx\langle\hat{H}_{\mathrm{B}}(t)\rangle_{\mathrm{h}}\,.
\end{SMEquations}

\noindent Now we consider the total work $W(t)=W_{\mathrm{on}}+W_{\mathrm{off}}(t)$, where we remind that (see Eq.~(46)) in the main text 
\begin{SMEquations}
W_{\mathrm{on}}=\langle\hat{H}_{\mathrm{C}}(0)\rangle=\frac{\Omega^2}{4\omega_0};\quad
W_{\mathrm{off}}(t)=-\langle\hat{H}_{\mathrm{C}}(t)\rangle\,.
\label{app:wtot}
\end{SMEquations}
The switching off part can be evaluated considering Eq.~(20) in the main text 
for $\langle\hat{H}_{\mathrm{C}}
(t)\rangle$:
\begin{SMEquations}
    \frac{\langle\hat{H}_{\mathrm{C}}(t)\rangle}{m}=\langle\hat{\dot{x}}^2(t)\rangle-\frac{1}{2}\frac{d^2}{dt^2}\langle\hat{x}^2(t)\rangle-\left(\omega_0^2+\frac{\Omega^2}{2}\right)\langle\hat{x}^2(t)\rangle
    \label{app:HCrewritten}
\end{SMEquations}
 evaluated with the homogeneous contributions (\ref{app:homo1}). Thus, 

\begin{SMEquations}
W_{\mathrm{off}}(t)\approx W_{\mathrm{off,h}}(t)=-\frac{\Omega^2}{4\omega_0}e^{-\omega_{\mathrm{D}} t}\cos^2(\Omega t)\,.
\label{app:woffh}
\end{SMEquations}
This  allows to write the final expression of the total work
\begin{SMEquations}
W(t)=\frac{\Omega^2}{4\omega_0}\left[1-e^{-\omega_{\mathrm{D}} t}\cos^2(\Omega t)\right]\,.
\end{SMEquations}
\begin{figure}[h!]
    \centering
    \includegraphics[width=0.9\textwidth]{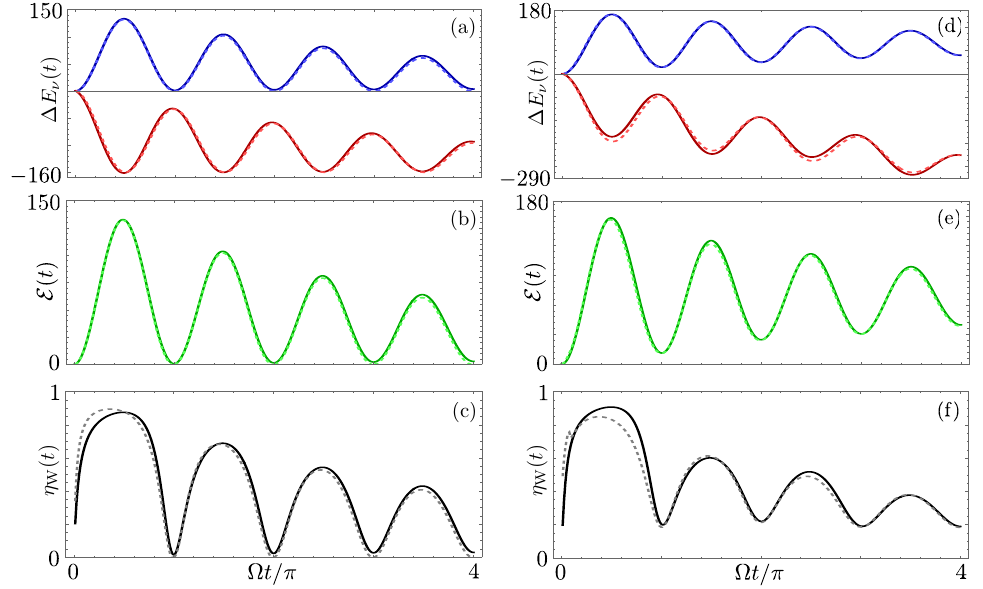}
    \caption{\textbf{Comparison between numerical and analytical results.} Panels (a-c) show respectively the plot of $\Delta E_{\nu}(t)$ with $\nu=\mathrm{B}$ (blue curve), $\nu=\mathrm{C}$ (red curve), of $\mathcal{E}(t)$ and of $\eta_{\mathrm{W}}(t)$ as a function of $\Omega t/\pi$ for the case $T=0.1\omega_0\approx4\cdot10^{-3}\Omega$. Panels (d-e) show the same quantities but for $T=250\omega_0\approx10\Omega$. In all panels the solid line represents the numerical results, while the dashed line represents the analytical ones. All energies are given in units $\omega_0$, other parameters are $\omega_{\mathrm{D}}=2 \omega_{0}$, $\gamma_{0}=300 \omega_{0}$.}
    \label{fig:figapp}
\end{figure}

\noindent These equations provide very accurate result, as shown in Fig.~\ref{fig:figapp} which shows a comparison between numerical (solid lines) and analytical (dashed lines) results. Both $\Delta E_{\mathrm{B}}(t)$ and $\Delta E_{\mathrm{C}}(t)$ (panel a), the ergotropy $\mathcal{E}(t)$ as well as $\eta_{\mathrm{W}}(t)$ display an excellent agreement.\\

\noindent{\bf \em -- High temperatures ($T\gg\omega_{\mathrm{D}}$)}\\

\noindent In this regime $\coth\left(\frac{\omega_{\mathrm{D}}}{2T}x \right)\to \frac{2T}{x\omega_{\mathrm{D}}}$ and the function $F(s)$ in Eq.~(\ref{app:F}) becomes  
\begin{SMEquations}
F(s)\approx\frac{\pi T}{\omega_{\mathrm{D}}}e^{-|s|}\,.
\end{SMEquations}
Inserting this expression in Eq.~(\ref{app:integralwithF}) with  $\chi(t)$ given in Eq.~(66) of the main text, 
after some manipulations it is possible to explicitly evaluate the integrals over $x_1$ and $x_2$. Below we quote the result to leading order in $\Omega\gg \omega_{\mathrm{D}},\omega_0$, valid for time $\Omega t$ up to the first local maxima of the QB which are at $\Omega t=n\pi/2$ with $n$ a small integer. We have
\begin{SMEqnarray}
&&\langle\hat{x}_{\mathrm{th}}^2(t)\rangle =\frac{2T}{m\Omega^2}\Big\{\Big[
1-e^{-\omega_{\mathrm{D}} t/2}\cos(\Omega t)\Big]-\frac{1}{2}e^{-\omega_{\mathrm{D}} t}\sin^2(\Omega t)+\frac{\Omega^2}{2\omega_0^2}\Big(1-e^{-2\omega_{\mathrm{D}}\omega_0^2 t/\Omega^2}\Big)
\Big\}\nonumber\\
&&\langle\hat{\dot x}_{\mathrm{th}}^2(t)\rangle =\frac{T}{m}\Big\{1-
e^{-\omega_{\mathrm{D}} t}\cos^2(\Omega t)\Big\}\,.
\label{app:thermal}
\end{SMEqnarray}

\noindent Notice that, in the regime of large $\Omega$ considered here, the variance $\langle\hat{\dot x}_{\mathrm{th}}^2(t)\rangle$  is always dominant with respect to $\langle\hat{x}_{\mathrm{th}}^2(t)\rangle$.
This implies $\langle\hat{H}_{\mathrm{B}}(t)\rangle_{\mathrm{th}}\approx \frac{m}{2}\langle\hat{\dot x}_{\mathrm{th}}^2(t)\rangle$, then considering also the homogeneous contribution, written in Eq.~(\ref{app:hb}), the total energy of the QB is 
\begin{SMEquations}
\langle\hat{H}_{\mathrm{B}}(t)\rangle=e^{-\omega_{\mathrm{D}} t}\sin^2(\Omega t)\left(\frac{\Omega^2}{4\omega_0}+\frac{T}{2}\right)
+\frac{T}{2}\Big[1-
e^{-\omega_{\mathrm{D}} t}\Big]\,.
\label{app:hbtot}
\end{SMEquations}
\!\!\! Concerning  the total work, written in Eq.~(\ref{app:wtot}), we see that the switching on work does not have thermal contributions, while the switching off ones is 
\begin{SMEquations}
W_{\mathrm{off}}(t)=W_{\mathrm{off,h}}(t)+W_{\mathrm{off, th}}(t)\,,
\end{SMEquations}
\!\!\!\! with the homogeneous part, written in Eq.~(\ref{app:woffh}) and  the thermal contribution $W_{\mathrm{off,th}}(t)=-\langle\hat{H}_{\mathrm{C}}(t)\rangle_{\mathrm{th}}$. This last term can be directly evaluated inserting in the expression (\ref{app:HCrewritten}) the above results for 
$\langle\hat{x}_{\mathrm{th}}^2(t)\rangle$ and $\langle\hat{\dot x}_{\mathrm{th}}^2(t)\rangle$. We then have, to leading order in $\Omega$
\begin{SMEquations}
\!\!\!\!\!W_{\mathrm{off,th}}(t)=\frac{T}{2}\left[e^{-\omega_{\mathrm{D}} t}\sin^2(\Omega t)+\frac{\Omega^2}{\omega_0^2}\left(\!1-e^{-\frac{2\omega_0^2\omega_{\mathrm{D}} t}{\Omega^2}}\right)\right]\,.
\end{SMEquations}

Putting all together we obtain the expression for the total work
\begin{SMEquations}
\!\!\!\!\!W(t)=-\Delta E_{\rm C}(t)=\frac{\Omega^2}{4\omega_0}\left[1-e^{-\omega_{\mathrm{D}} t}\cos^2(\Omega t)\right]+\frac{T}{2}\left[e^{-\omega_{\mathrm{D}} t}\sin^2(\Omega t)+\frac{\Omega^2}{\omega_0^2}\left(\!1-e^{-\frac{2\omega_0^2\omega_{\mathrm{D}} t}{\Omega^2}}\right)\right]\,.
\end{SMEquations}

\noindent We close by commenting on the ergotropy $\mathcal{E}(t)$. Differently from the low temperature case, at very high $T\gg \Omega$ the contribution of the covariance matrix becomes relevant and cannot be disregarded. The analytical evaluation of ${\rm det}\left[\sigma_{\mathrm{B,h}}(t)+\sigma_{\mathrm{B,th}}(t)\right]$ can be directly obtained using the final expressions for the homogeneous variances -- Eq.(\ref{app:homo1}) -- and for the thermal ones -- Eq.(\ref{app:thermal}). The final expressions are too lengthy to be reported here.\\ 

\noindent Panels (d-f) of Fig.~\ref{fig:figapp} confirm that also in the high temperature case the accordance between numerical and analytical results, which is excellent for both $\Delta E_{\mathrm{B}}(t)$ and for the ergotropy and very good for $\Delta E_{\mathrm{C}}(t)$ and the efficiency $\eta_{\mathrm{W}}(t)$.\\

\centerline{\bf b. Long times}

This part is devoted to  determine the long time ($t\rightarrow +\infty$) behavior of the figures of merit. First, we remind that  at long times the homogeneous contributions die out and only the thermal ones survive. In addition, as shown in Eq.~(81) of the main text, 
we have $\langle \hat{x}_{\mathrm{th}}(+\infty)\hat{\dot x}_{\mathrm{th}}(+\infty)\rangle=0$, then  only $\langle\hat{x}^{2}_{\mathrm{th}}(+\infty)\rangle=\langle \hat{x}^{2}(+\infty)\rangle$ and $\langle \hat{\dot x}^{2}_{\mathrm{th}}(+\infty)\rangle=\langle \hat{\dot x}^{2}(+\infty)\rangle$ survive. This implies that the ergotropy, see Eq.~(17) in the main text 
, is 
\begin{SMEquations}
\mathcal{E}(+\infty)=\frac{\omega_0}{2}\left[\sqrt{m\omega_0\langle \hat{x}^{2}(+\infty)\rangle}-\sqrt{m\langle \hat{\dot x}^{2}(+\infty)\rangle/\omega_0} 
\right]^2\,,
\end{SMEquations}
while for the total work we have
\begin{SMEqnarray}
W(\infty)&=&\frac{\Omega^2}{4\omega_0}-\langle\hat{H}_{\mathrm{C}}(\infty)\rangle\\
\langle\hat{H}_{\mathrm{C}}(\infty)\rangle&=&m \langle \hat{\dot x}^{2}(+\infty)\rangle-m\left(\omega_0^2+\frac{\Omega^2}{2}\right)\langle\hat{x}^2(\infty)\rangle\,.\nonumber
\label{app:Winfty}
\end{SMEqnarray}
In order to evaluate the above quantities we  use  the general results derived in Ref.~\cite{Weiss_book} for a Drude spectral density. We have 
\begin{SMEqnarray}
\langle \hat{x}^{2}(+\infty)\rangle&=&\frac{T}{m \omega^{2}_{0}}-\frac{1}{m \pi} \sum^{3}_{j=1} \chi_{j}\psi\left(1+\frac{\lambda_{j}}{2 \pi T}\right), \qquad\langle \hat{\dot x}^{2}(+\infty)\rangle=\omega^{2}_{0}\, \langle \hat{x}^{2}_{\mathrm{th}}(+\infty)\rangle+K
\label{app:xpunto2}\\
\stepcounter{SMEquation}
K&=&\frac{\Omega^2}{m\pi}\sum^{3}_{j=1}k_{j} \psi\left(1+\frac{\lambda_{j}}{2 \pi T}\right)\,,
\label{app:kappa}
\end{SMEqnarray}
with
\begin{SMEquations}
k_{j}=\frac{\lambda_{j}}{\prod_{j'\neq j}\left(\lambda_{j'}-\lambda_{j} \right)}
\end{SMEquations}

\noindent and where $\psi(z)$ is the digamma function.\\
\noindent Our task will be to study these relations in the strong underdamped regime ($\Omega\gg \omega_0,\omega_{\mathrm{D}}$). Here, the roots $\lambda_j$ and the corresponding weight $\chi_j$ were obtained in Eqs.~(63) -
~(65) of the main text.
We will confine the discussion below to the two limiting cases of low or high temperatures.\\

\noindent-- {\bf \em Low temperatures ($T\ll\lambda_j$)}

\noindent In this case one has $\psi\left(1+\frac{\beta \lambda_{j}}{2 \pi}\right)=\ln\left(\frac{\beta \lambda_{j}}{2 \pi} \right)+O(T/\lambda_j)$, this leads to the leading contribution
\begin{SMEquations}
\langle \hat{x}^{2}(+\infty)\rangle \approx -\frac{1}{\pi m }\sum^{3}_{j=1}\chi_{j} \ln \left(\frac{\lambda_{j}}{2 \pi T} \right)=-\frac{1}{\pi m}\sum^{3}_{j=1}\chi_{j} \ln\left(\lambda_{j}\right)\,,
\end{SMEquations}
\!\!\!\! where in the last equation we use $\sum^{3}_{j=1}\chi_{j}\!=\!0$.
Inserting the expressions for $\lambda_j$ and $\chi_j$
we obtain the leading behavior in $\Omega$ for $T\to 0$
\begin{SMEquations}
\langle \hat{x}^{2}(+\infty)\rangle\approx\frac{1}{2m \Omega}.
\label{app:x2finale}
\end{SMEquations}
To determine  $\langle \hat{\dot x}^{2}(+\infty)\rangle$ we start by considering the contribution $K$ in Eq.~(\ref{app:kappa}) for $T\to 0$. We have 
\begin{SMEquations}
K\approx\frac{\Omega^2}{\pi m}\sum^{3}_{j=1}k_{j}\ln \left(\frac{\lambda_{j}}{2 \pi T} \right)
\end{SMEquations}
Now we use the constraint $\sum^{3}_{j=1}k_{j}=0$. This follows from the equality 
\begin{SMEquations}
    k_j=-\chi_j+\frac{\omega_{\mathrm{D}}}{\prod_{j'\neq j}\left(\lambda_{j'}-\lambda_{j} \right)}\,,
\end{SMEquations}
recalling $\sum^{3}_{j=1}\chi_{j}=0$ and observing that $\sum^{3}_{j=1}\frac{1}{\prod_{j'\neq j}\left(\lambda_{j'}-\lambda_{j} \right)}=0$. We are then left with the following expression
\begin{SMEquations}
K\approx\frac{\Omega^2}{\pi m}\sum^{3}_{j=1}k_{j}\ln \left(\lambda_{j}\right)\,.
\end{SMEquations}
Inserting the leading behavior in $\Omega$ of $\lambda_j$ and $\chi_j$
we have 
\begin{SMEquations}
K\approx \frac{\Omega}{2m}
\label{app:kfinale}
\end{SMEquations}
This implies, using Eqs.~(\ref{app:xpunto2}),\,(\ref{app:x2finale}) and (\ref{app:kfinale}), that the leading behavior in $\Omega$  for $\langle \hat{\dot x}^{2}(+\infty)\rangle$ is 
\begin{SMEquations}
\langle \hat{\dot x}^{2}(+\infty)\rangle \approx \frac{\Omega}{2m}.
\end{SMEquations}

\noindent Regarding the different figure of merit we then have  
\begin{SMEquations}
\langle \hat{H}_{\mathrm{B}}(+\infty)\rangle\approx \frac{\Omega}{4}\,,\quad\langle \mathcal{E}(+\infty)\rangle\approx\frac{\Omega}{4}
\end{SMEquations}\\

\noindent and therefore $\eta_{\mathrm{B}}(+\infty)\approx 1$.  Inspecting Eq.~(\ref{app:Winfty}) we see that the work done to switch on the interaction $W_{\mathrm{on}}=\Omega^2/(4\omega_0)$ dominates with respect to the work required to switch off the coupling.
Indeed, the latter is (see Eq~(\ref{app:Winfty})) $W_{\mathrm{off}}=-\langle\hat{H}_{\mathrm{C}}(\infty)\rangle\approx-\Omega/4$. Then the total work is 
\begin{SMEquations}
W(+\infty)\approx \frac{\Omega^{2}}{4 \omega_{0}}
\end{SMEquations}
which leads to a very small efficiency  
\begin{SMEquations}
\eta_{\mathrm{W}}(+\infty)\approx \frac{\omega_{0}}{\Omega}\to 0.
\end{SMEquations}\\
\noindent-- {\bf \em High  temperatures (\,$T\gg\lambda_j$)}
\noindent In this regime  we consider the high temperature expansion of all terms containing the psi function:
$\psi\left(1+\frac{\beta \lambda_{j}}{2 \pi}\right)=\psi(1)+O\left(\frac{\lambda_{j}}{T}\right)$.
Inserting this behavior in Eq.~(\ref{app:kappa}) we can see that the $K$ term is not dominant ($K\propto 1/T$). Indeed, we have  
\begin{SMEquations}
\langle \hat{x}^{2}(+\infty)\rangle=\frac{T}{m \omega^{2}_{0}}\,[1+O(\Omega/T)],\quad 
\langle \hat{\dot x}^{2}(+\infty)\rangle=\frac{T}{m}\,[1+O(\Omega/T)].
\label{app:finalalteT}
\end{SMEquations}

This result is no more then the energy equipartition. In this regime one then has
\begin{SMEquations}
\langle \hat{H}_{\mathrm{B}}(+\infty)\rangle=\,T\,\qquad
 \mathcal{E}(+\infty)\propto 1/T
\end{SMEquations}
and therefore $\eta_{\mathrm{B}}(+\infty)\to 0$.\\
\noindent We conclude addressing the total work $W(t)$. Inserting Eq.~(\ref{app:finalalteT}) in Eq.~(\ref{app:Winfty}) we obtain
\begin{SMEquations}
W(+\infty)=\frac{T\,\Omega^{2}}{2\omega^{2}_{0}}[1+O(\Omega/T)]
\end{SMEquations}

which leads to $\eta_{\mathrm{W}}(+\infty)\to 0$.

\section*{Supplementary Note 3: Details on the spectral decomposition of the reservoir and coupling energies}
\label{App:app5}
Let us begin observing that $\hat{H}_{\mathrm{C}}$ in Eq.~(4) in the main text can be rewritten as

\begin{SMEquations}
\hat{H}_{\mathrm{C}}=\frac{1}{2}\sum_{k}\left[-c_k\left(\hat{x}\hat{x}_k+\hat{x}_k\hat{x}\right)+\frac{c_k^2}{m_k\omega_k^2}\hat{x}^2\right]\equiv\sum_{k}\hat{H}_{\mathrm{C}}^{(k)}\,,\label{eq:1A}
\end{SMEquations}
whence
\begin{SMEquations}
\langle\hat{H}_{\mathrm{C}}^{(k)}(t)\rangle=-c_k\langle\hat{x}(t)\hat{x}_k(t)\rangle_{\mathrm{s}}+\frac{c_k^2}{2m_k\omega_k^2}\langle\hat{x}^2(t)\rangle\,,\label{eq:1AA}
\end{SMEquations}
where $\langle\hat{A}\hat{B}\rangle_{\mathrm{s}}=\frac{1}{2}\langle\{\hat{A},\hat{B}\}\rangle$ is the symmetrized average. Let us focus on the first term $\langle\hat{x}(t),\hat{x}_k(t)\rangle_{\mathrm{s}}$, since the quantity $\langle\hat{x}^2(t)\rangle$ has already been evaluated -- see Eqs.~(43) and~(44) in the main text. Also in the case of this correlation function, the homogeneous and thermal contributions are separated, leading to
\begin{SMEquations}
\langle\hat{x}(t),\hat{x}_k(t)\rangle_{\mathrm{s}}=\langle\hat{x}_{\mathrm{h}}(t),\hat{x}_{k,\mathrm{h}}(t)\rangle_{\mathrm{s}}+\langle\hat{x}_{\mathrm{th}}(t),\hat{x}_{k,\mathrm{th}}(t)\rangle_{\mathrm{s}}\,,\label{eq:2A}
\end{SMEquations}
where $\hat{x}_{\mathrm{h}}(t)$ and $\hat{x}_{\mathrm{th}}(t)$ have been given in Eqs.~(13) and~(15) of the main text, while from Eq.~(8) we have
\begin{SMEquations}
    \hat{x}_{k,\mathrm{th}}(t)=\hat{\xi}_k(t)+\frac{c_k}{m_k\omega_k}\int_{0}^{t}dt'\sin\left[\omega_k\left(t-t'\right)\right]\hat{x}_{\mathrm{th}}(t')\,,\label{eq:3A}
\end{SMEquations}
with
\begin{SMEquations}
\hat{\xi}_k(t)=\hat{x}_k(0)\cos\left(\omega_k t\right)+\frac{\hat{p}_k(0)}{m_k\omega_k}\sin\left(\omega_k t\right),\quad{\rm and}\quad\hat{x}_{k,\mathrm{h}}=\frac{c_k}{m_k\omega_k}\int_{0}^{t}dt'\sin\left[\omega_k\left(t-t'\right)\right]\hat{x}_{\mathrm{h}}(t')\,.\label{eq:5A}
\end{SMEquations}

We recall here that the QB is initialized from its ground state, with
\begin{SMEquations}
\langle\hat{x}^2(0)\rangle=\frac{1}{2m\omega_0};\quad \langle\hat{\dot{x}}^2(0)\rangle=\frac{\omega_0}{2m};\quad \langle\hat{x}(0)\hat{\dot{x}}(0)\rangle_{\mathrm{s}}=0\,,\label{eq:6AA}
\end{SMEquations}
while the reservoir has thermal initial conditions
\begin{SMEqnarray}
\langle\hat{x}_k(0)\hat{x}_{k'}(0)\rangle=\frac{\delta_{k,k'}}{2m_k\omega_k}\coth\left(\frac{\beta\omega_k}{2}\right);\quad \langle\hat{p}_k(0)\hat{p}_{k'}(0)\rangle=\frac{\delta_{k,k'}}{2}m_k\omega_k\coth\left(\frac{\beta\omega_k}{2}\right);\quad \langle\hat{x}_k(0)\hat{p}_{k'}(0)\rangle_{\mathrm{s}}=0\,.\label{eq:6A}
\end{SMEqnarray}
To evaluate the homogeneous term $\langle\hat{x}_{\mathrm{h}}(t)\hat{x}_{k,\mathrm{h}}(t)\rangle_{\mathrm{s}}$ in Eq.~(\ref{eq:2A}) using Eq.~(\ref{eq:5A}) one needs the correlator $\langle\hat{x}_{\mathrm{h}}(t)\hat{x}_{\mathrm{h}}(t')\rangle_{\mathrm{s}}$, which can be easily obtained using Eq.~(13) in the main text with the initial conditions (\ref{eq:6AA}). One has
\begin{SMEquations}
    \langle\hat{x}_{\mathrm{h}}(t)\hat{x}_{\mathrm{h}}(t')\rangle_{\mathrm{s}}=\frac{1}{2m\omega_0}\left[\omega_0^2\chi(t)\chi(t')+\dot{\chi}(t)\dot{\chi}(t')\right]\label{eq:6AAA}\,.
\end{SMEquations}
Inserting this expression in Eq.~(\ref{eq:2A}) one obtains
\begin{SMEquations}
\langle\hat{x}_{\mathrm{h}}(t)\hat{x}_{\mathrm{h}}(t')\rangle_{\mathrm{s}}=\frac{\omega_0}{2m}\frac{c_k}{m_k\omega_k}\sum_{l,j}\chi_l\chi_j\left(1+\frac{\lambda_l\lambda_j}{\omega_0^2}\right)\frac{e^{-(\lambda_l+\lambda_j)t}}{2i}\left[\frac{e^{(\lambda_j+i\omega_k)t}-1}{\lambda_j+i\omega_k}-\frac{e^{(\lambda_j-i\omega_k)t}-1}{\lambda_j-i\omega_k}\right]\,.\label{eq:7A}
\end{SMEquations}
Let us now turn to the evaluation of the term $\langle\hat{x}_{\mathrm{th}}(t)\hat{x}_{k,\mathrm{th}}(t)\rangle_{\mathrm{s}}$ in Eq.~(\ref{eq:2A}). Inspecting Eq.~(\ref{eq:3A}) one figures out that it consists of two contributions
\begin{SMEquations}
\langle\hat{x}_{\mathrm{th}}(t)\hat{x}_{k,\mathrm{th}}(t)\rangle_{\mathrm{s}}=\langle\hat{x}_{\mathrm{th}}(t)\hat{\xi}_{k}(t)\rangle_{\mathrm{s}}+\frac{c_k}{m_k\omega_k}\int_{0}^{t}dt'\sin\left[\omega_k(t-t')\right]\langle\hat{x}_{\mathrm{th}}(t)\hat{x}_{\mathrm{th}}(t')\rangle_{\mathrm{s}}\,.\label{eq:8A}
\end{SMEquations}
Starting from the first term on the right hand side and substituting the solution for $\hat{x}_{\mathrm{th}}(t)$ -- see Eq.~(15) in the main text -- one has
\begin{SMEquations}
\langle\hat{x}_{\mathrm{th}}(t)\hat{\xi}_{k}(t)\rangle_{\mathrm{s}}=\int_{0}^{t}\frac{dt'}{m}\chi(t-t')\sum_{k'}c_{k'}\langle\hat{\xi}_{k'}(t')\hat{\xi}_{k}(t)\rangle_{\mathrm{s}}\,.\label{eq:9A}
\end{SMEquations}
Taking into account Eq.~(\ref{eq:5A}) and the initial conditions (\ref{eq:6A}) one finds
\begin{SMEquations}
\langle\hat{\xi}_{k'}(t')\hat{\xi}_{k}(t)\rangle_{\mathrm{s}}=\frac{\delta_{k,k'}}{2m_k\omega_k}\coth\left(\frac{\beta\omega_k}{2}\right)\cos\left[\omega_k\left(t-t'\right)\right]\,.\label{eq:10A}
\end{SMEquations}
Plugging Eq.~(\ref{eq:10A}) into Eq.~(\ref{eq:9A}) with  the general expression $\chi(t)=\sum_{j}\chi_j e^{-\lambda_j t}$ after some algebraic steps we have
\begin{SMEquations}
\langle\hat{x}_{\mathrm{th}}(t)\hat{\xi}_{k}(t)\rangle_{\mathrm{s}}=\frac{1}{m}\frac{c_k}{4m_k\omega_k}\coth\left(\frac{\beta\omega_k}{2}\right)\sum_{j}\chi_j\left[\frac{1-e^{-(\lambda_j-i\omega_k)t}}{\lambda_j-i\omega_k}+\frac{1-e^{-(\lambda_j+i\omega_k)t}}{\lambda_j+i\omega_k}\right]\,.\label{eq:11A}
\end{SMEquations}
As for the last term in Eq.~(\ref{eq:8A}) we need the following correlator
\begin{SMEquations}
\langle\hat{x}_{\mathrm{th}}(t)\hat{x}_{\mathrm{th}}(t')\rangle_{\mathrm{s}}=\frac{1}{m^2}\int_{0}^{t}d\tau_1\chi(t-\tau_1)\int_{0}^{t}d\tau_2\chi(t'-\tau_2)\mathcal{L}_{\mathrm{s}}(\tau_1-\tau_2)\,,\label{eq:12A}
\end{SMEquations}
\!\!\!\! where $\mathcal{L}_\mathrm{s}(\tau)=\langle\hat{\xi}(\tau)\hat{\xi}(0)\rangle_{\mathrm{s}}$ is given in Eq.~(\ref{app:Lcompact}) and where we have employed Eq.~(15) in the main text. Using again the general form of $\chi(t)$ quoted above and performing the integrations in Eq.~(\ref{eq:12A}) we get

\begin{SMEquations}
\langle\hat{x}_{\mathrm{th}}(t)\hat{x}_{\mathrm{th}}(t')\rangle_{\mathrm{s}}=\frac{1}{m^2}\sum_{l,j}\chi_l\chi_j\int_{-\infty}^{\infty}\frac{d\omega}{2\pi}J(\omega)\cdot\coth\left(\frac{\beta\omega}{2}\right)\frac{e^{i\omega t}-e^{-\lambda_l t}}{\lambda_l+i\omega}\frac{e^{-i\omega t'}-e^{-\lambda_j t'}}{\lambda_j-i\omega}\,.\label{eq:14A}
\end{SMEquations}
\!\!\!\! This last correlator must be inserted back into the integral to the right hand side of Eq.~(\ref{eq:8A}). After some algebraic rearrangement, its last term finally reads

\begin{SMEqnarray}
&&\frac{1}{2m^2}\frac{c_k}{m_k\omega_k}\sum_{l,j}\chi_l\chi_j\int_{-\infty}^{\infty}\frac{d\omega}{2\pi}J(\omega)\coth\left(\frac{\beta\omega}{2}\right)\frac{e^{i\omega t}-e^{-\lambda_l t}}{\left(\lambda_l+i\omega\right)\left(\lambda_j-i\omega\right)}\left\{\frac{e^{-i\omega_k t}-e^{-i\omega t}}{\omega-\omega_k}-\frac{e^{i\omega_k t}-e^{-i\omega t}}{\omega+\omega_k}+\right.\nonumber\\
&&\left.+i\frac{e^{i\omega_k t}-e^{-\lambda_j t}}{\lambda_j+i\omega_k}-i\frac{e^{-i\omega_k t}-e^{-\lambda_j t}}{\lambda_j-i\omega_k}\right\}\,.\label{eq:15A}
\end{SMEqnarray}

To summarize, the term $\langle\hat{x}(t)\hat{x}_k(t)\rangle_{\mathrm{s}}$ is given by the sum of Eqs.~(\ref{eq:7A}),~(\ref{eq:11A}) and~(\ref{eq:15A}). This expression is to be inserted in Eq.~(\ref{eq:1AA}) to finally obtain $\langle\hat{H}_\mathrm{C}^{(k)}(t)\rangle$ and thus $\Delta E_{\mathrm{C}}(t,\omega_k)$ in Eq.~(24) of the main text.\\

\noindent Let us now evaluate the $k$--th contribution to the reservoir energy
\begin{SMEquations}
\langle \hat{H}_{\mathrm{R}}^{(k)}\rangle=\frac{1}{2}m_k\langle\hat{\dot{x}}_k^2(t)\rangle+\frac{1}{2}m_k\omega_k^2\langle\hat{x}_k^2(t)\rangle\,,\label{eq:17A}
\end{SMEquations}
again given by homogeneous and thermal contributions
\begin{SMEquations}
\langle\hat{H}_{\mathrm{R}}^{(k)}(t)\rangle=\langle\hat{H}_{\mathrm{R}}^{(k)}(t)\rangle_{\mathrm{h}}+\langle\hat{H}_{\mathrm{R}}^{(k)}(t)\rangle_{\mathrm{th}}\label{eq:18A}\,,   
\end{SMEquations}
where
\begin{SMEqnarray}
&&\langle\hat{H}_{\mathrm{R}}^{(k)}\rangle_{\mathrm{h}}=\frac{1}{2}m_k\langle\hat{\dot{x}}_{k,\mathrm{h}}^2(t)\rangle+\frac{1}{2}m_k\omega_k^2\langle\hat{x}_{k,\mathrm{h}}^2(t)\rangle\,,\label{eq:19A1}\nonumber\\
&&\langle\hat{H}_{\mathrm{R}}^{(k)}\rangle_{\mathrm{th}}=\frac{1}{2}m_k\langle\hat{\dot{x}}_{k,\mathrm{th}}^2(t)\rangle+\frac{1}{2}m_k\omega_k^2\langle\hat{x}_{k,\mathrm{th}}^2(t)\rangle\,.\label{eq:19A2}
\end{SMEqnarray}
Begin with $\langle\hat{H}_{\mathrm{R}}^{(k)}(t)\rangle_{\mathrm{h}}$, inserting Eq.~(\ref{eq:5A}) and its derivative to obtain

\begin{SMEquations}    \begin{pmatrix}\langle\hat{x}_{k,\mathrm{h}}^2(t)\rangle\\\\\langle\hat{\dot{x}}_{k,\mathrm{h}}^2(t)\rangle\end{pmatrix}=\left(\frac{c_k}{m_k\omega_k}\right)^2\int_{0}^{t}d\tau_1\int_{0}^{t}d\tau_2\langle\hat{x}_{\mathrm{h}}(\tau_1)\hat{x}_{\mathrm{h}}(\tau_2)\rangle\begin{pmatrix}\sin\left[\omega_k\left(t-\tau_1\right)\right]\sin\left[\omega_k\left(t-\tau_2\right)\right]\\\\\omega_k^2\cos\left[\omega_k\left(t-\tau_1\right)\right]\cos\left[\omega_k\left(t-\tau_2\right)\right]\end{pmatrix}\,.\label{eq:20A}
\end{SMEquations}

Inserting the solution in Eq.~(\ref{eq:6AAA}) in Eq.~(\ref{eq:20A}), after some steps one finds

\begin{SMEquations}
\langle\hat{H}_{\mathrm{R}}^{(k)}(t)\rangle_{\mathrm{h}}=\frac{c_k^2}{4m_k}\frac{\omega_0}{m}\sum_{l,j}\chi_l\chi_j\left(1+\frac{\lambda_l\lambda_j}{\omega_0^2}\right)\left[\frac{1-e^{-(\lambda_l+i\omega_k)t}}{\lambda_l+i\omega_k}\right]\left[\frac{1-e^{-(\lambda_j-i\omega_k)t}}{\lambda_j-i\omega_k}\right]\,.\label{eq:21A}
\end{SMEquations}

We are left with the evaluation of $\langle\hat{H}_{\mathrm{R}}^{(k)}(t)\rangle_{\mathrm{th}}$ given in Eq.~(\ref{eq:19A2}). To this end we need to rewrite $\langle\hat{\dot{x}}_{k,\mathrm{th}}^2(t)\rangle$ and $\langle\hat{x}_{k,\mathrm{th}}^2(t)\rangle$ using Eqs.~(\ref{eq:3A})--~(\ref{eq:5A}). Let us begin with $\langle\hat{x}_{k,\mathrm{th}}^2(t)\rangle$ which, after inserting Eq.~(\ref{eq:3A}), becomes

\begin{SMEqnarray}
&&\langle\hat{x}_{k,\mathrm{th}}^2(t)\rangle=\langle\hat{\xi}_k^2(t)\rangle+\frac{2c_k}{m_k\omega_k}\int_{0}^{t}dt'\sin\left[\omega_k\left(t-t'\right)\right]\langle\hat{\xi}_k(t)\hat{x}_{\mathrm{th}}(t')\rangle_{\mathrm{s}}+\left(\frac{c_k}{m_k\omega_k}\right)^2\int_{0}^{t}d\tau_1\int_{0}^{t}d\tau_2\sin\left[\omega_k\left(t-\tau_1\right)\right]\times\nonumber\\
&&\times\sin\left[\omega_k\left(t-\tau_2\right)\right]\langle\hat{x}_{\mathrm{th}}(\tau_1)\hat{x}_{\mathrm{th}}(\tau_2)\rangle_{\mathrm{s}}\,.\label{eq:22A}
\end{SMEqnarray}

In analogy, concerning $\langle\hat{\dot{x}}_{k,\mathrm{th}}^2(t)\rangle$ we have
\begin{SMEqnarray}
&&\langle\hat{\dot{x}}_{k,\mathrm{th}}^2(t)\rangle=\langle\hat{\dot{\xi}}_k^2(t)\rangle+\frac{2c_k}{m_k}\int_{0}^{t}dt'\cos\left[\omega_k\left(t-t'\right)\right]\langle\hat{\dot{\xi}}_k(t)\hat{x}_{\mathrm{th}}(t')\rangle_{\mathrm{s}}+\left(\frac{c_k}{m_k}\right)^2\int_{0}^{t}d\tau_1\int_{0}^{t}d\tau_2\cos\left[\omega_k\left(t-\tau_1\right)\right]\times\nonumber\\
&&\times\cos\left[\omega_k\left(t-\tau_2\right)\right]\langle\hat{x}_{\mathrm{th}}(\tau_1)\hat{x}_{\mathrm{th}}(\tau_2)\rangle_{\mathrm{s}}\,.\label{eq:23A}
\end{SMEqnarray}

\noindent After inserting Eqs.~(\ref{eq:22A}) and~(\ref{eq:23A}), Eq.~(\ref{eq:19A2}) decomposes as
\begin{SMEquations}
\langle\hat{H}_{\mathrm{R}}^{(k)}(t)\rangle_{\mathrm{th}}=I_0^{(k)}+I_1^{(k)}(t)+I_2^{(k)}(t)\,,\label{eq:24A}
\end{SMEquations}
where
\begin{SMEquations}
I_0^{(k)}=\frac{1}{2}m_k\langle\hat{\dot{\xi}}_k^2(t)\rangle+\frac{1}{2}m_k\omega_k^2\langle\hat{\xi}_k^2(t)\rangle\,,\label{eq:25A}
\end{SMEquations}
\begin{SMEquations}
I_1^{(k)}(t)=c_k\int_{0}^{t}dt'\left\{\cos\left[\omega_k\left(t-t'\right)\right]\langle\hat{\dot{\xi}}_k(t)\hat{x}_{\mathrm{th}}(t')\rangle_{\mathrm{s}}+\omega_k\sin\left[\omega_k\left(t-t'\right)\right]\langle\hat{\xi}_k(t)\hat{x}_{\mathrm{th}}(t')\rangle_{\mathrm{s}}\right\}\,,\label{eq:26A}
\end{SMEquations}
and
\begin{SMEquations}
I_2^{(k)}(t)=\frac{c_k^2}{2m_k}\int_{0}^{t}d\tau_1\int_{0}^{t}d\tau_2\cos\left[\omega_k\left(\tau_1-\tau_2\right)\right]\langle\hat{x}_{\mathrm{th}}(\tau_1)\hat{x}_{\mathrm{th}}(\tau_2)\rangle_{\mathrm{s}}\,.\label{eq:27A}
\end{SMEquations}
Substituting Eq.~(\ref{eq:5A}) in Eq.~(\ref{eq:25A}) and keeping in mind the initial conditions (\ref{eq:6A}) we get
\begin{SMEquations}
I_0^{(k)}=\frac{\omega_k}{2}\coth\left(\frac{\beta\omega_k}{2}\right)\,,\label{eq:28A}
\end{SMEquations}
which is independent of $t$ as one should expect being independent of $c_k$. To evaluate $I_1^{(k)}(t)$ in Eq~(\ref{eq:26A}) we need $\langle\hat{\xi}_{k}(t)\hat{x}_{\mathrm{th}}(t')\rangle_{\mathrm{s}}$, which can be obtained recalling Eq.~(15) in the main text and the correlator in Eq.~(\ref{eq:10A}). Doing so one gets
\begin{SMEquations}
\langle\hat{\xi}_k(t)\hat{x}_{\mathrm{th}}(t')\rangle_{\mathrm{s}}=\frac{c_k}{2m_k\omega_k}\coth\left(\frac{\beta\omega_k}{2}\right)\int_{0}^{t'}\frac{d\tau}{m}\chi(t'-\tau)\cos\left[\omega_k(t-\tau)\right]\,.\label{eq:29A}
\end{SMEquations}
Note that to obtain $\langle\hat{\dot{\xi}}_{k}(t)\hat{x}_{\mathrm{th}}(t')\rangle_{\mathrm{s}}$ one simply needs to take the derivative of Eq.~(\ref{eq:29A}) with respect to $t$. Inserting these quantities into Eq.~(\ref{eq:26A}) we obtain
\begin{SMEquations}
I_1^{(k)}(t)=\frac{c_k^2}{4mm_k}\coth\left(\frac{\beta\omega_k}{2}\right)\sum_{j}\chi_j\left\{-\frac{2\omega_k t}{\lambda_j^2+\omega_k^2}+i\frac{1-e^{-(\lambda_j+i\omega_k)t}}{\left(\lambda_j+i\omega_k\right)^2}-i\frac{1-e^{-(\lambda_j-i\omega_k)t}}{\left(\lambda_j-i\omega_k\right)^2}\right\}\,.\label{eq:30A}
\end{SMEquations}
Note here the presence of a secular term, which in a minute we will show to perfectly cancel out for $t\to\infty$ with another secular term appearing in $I_2^{(k)}(t)$. As for this last term, it is easy to evaluate inserting Eq.~(\ref{eq:14A}) for the correlator $\langle\hat{x}_{\mathrm{th}}(\tau_1)\hat{x}_{\mathrm{th}}(\tau_2)\rangle_{\mathrm{s}}$. After a bit of algebra we get

\begin{SMEqnarray}
&&I_2^{(k)}(t)=\frac{c_k^2}{2m_k}\frac{1}{m^2}\sum_{l,j}\int_{-\infty}^{\infty}\frac{d\omega}{2\pi}\coth\left(\frac{\beta\omega}{2}\right)\frac{J(\omega)}{\left(\lambda_l+i\omega\right)\left(\lambda_j-i\omega\right)}\left[\frac{e^{-i(\omega-\omega_k)t}-1}{\omega-\omega_k}-i\frac{e^{-\lambda_j-i\omega_k)t}-1}{\lambda_j-i\omega_k}\right]\times\nonumber\\
&&\times\left[\frac{e^{i(\omega-\omega_k)t}-1}{\omega-\omega_k}+i\frac{e^{-\lambda_l+i\omega_k)t}-1}{\lambda_l+i\omega_k}\right]\,.\label{eq:31A}
\end{SMEqnarray}

To summarize, $\langle\hat{H}_{\mathrm{R}}^{(k)}(t)\rangle$ is given by the sum of the terms in Eq.~(\ref{eq:21A}),~(\ref{eq:28A}),~(\ref{eq:30A}) and~(\ref{eq:31A}).

To close this Section let us show that for $t\to\infty$ a secular term emerges in $I_2^{(k)}(t)$ which exactly cancels out the secular term in $I_1^{(k)}(t)$. To extract these terms we  consider $\lim_{t\to\infty}\dot{I}_{1,2}^{(k)}(t)$, the sum of them should cancel out. We have

\begin{SMEquations}
\lim_{t\to\infty}\dot{I}_1^{(k)}(t)=-\frac{c^2_k\omega_k}{2mm_k}\coth\left(\frac{\beta\omega_k}{2}\right)\sum_{j}\frac{\chi_j}{\lambda_j^2+\omega_k^2}\,.\label{eq:32A} 
\end{SMEquations}
Notice that

\begin{SMEquations}
\sum_{j}\frac{\chi_j}{\lambda_j-i\omega}=\int_{0}^{\infty}dt\chi(t)e^{i\omega t}=\tilde{\chi}(\lambda=-i\omega)\,,\label{eq:33A}
\end{SMEquations}
\!\!\!\! and that $\tilde{\chi}(\lambda=-i\omega)$ turns out to be the Fourier transform of the retarded response function $\chi_{\mathrm{r}}(t)=\theta(t)\chi(t)$: $\tilde{\chi}_{\mathrm{r}}(\omega)=\tilde{\chi}(\lambda=-i\omega)$ -- see Eq.~(14) in the main text. Therefore, Eq.~(\ref{eq:32A}) becomes
\begin{SMEquations}
\lim_{t\to\infty}\dot{I}_1^{(k)}(t)=-\frac{c^2_k}{2mm_k}\coth\left(\frac{\beta\omega_k}{2}\right)\tilde{\chi}''_{\mathrm{r}}(\omega_k)\,,\label{eq:34A} 
\end{SMEquations}
where $\tilde{\chi}''_{\mathrm{r}}(\omega_k)$ denotes the imaginary part of $\tilde{\chi}_{\mathrm{r}}(\omega_k)$. Consider now 

\begin{SMEqnarray}
&&\lim_{t\to\infty}\dot{I}_2^{(k)}(t)=\frac{c_k^2}{2m_m}\frac{1}{m^2}\sum_{l,j}\chi_l\chi_j\int_{-\infty}^{\infty}\frac{d\omega}{2\pi}\coth\left(\frac{\beta\omega}{2}\right)\frac{J(\omega)}{\left(\lambda_l+i\omega\right)\left(\lambda_j-i\omega\right)}\left\{\frac{2\sin\left[\left(\omega-\omega_k\right)t\right]}{\omega-\omega_k}\right.\nonumber\\
&&\left.-\left[\frac{e^{-i(\omega-\omega_k)t}}{\lambda_l+i\omega_k}+\frac{e^{i(\omega-\omega_k)t}}{\lambda_j-i\omega_k}\right]\right\}\,,
\end{SMEqnarray}
\!\!\!\! and introduce the auxiliary variable $x=(\omega-\omega_k)t$ and let $t\to\infty$ in the integral of the above expression. The only surviving term reads
\begin{SMEquations}
\lim_{t\to\infty}\dot{I}_2^{(k)}(t)=\frac{c_k^2}{2m_k}\frac{1}{m^2}\coth\left(\frac{\beta\omega_k}{2}\right)J(\omega_k)\sum_{l,j}\frac{\chi_l}{\lambda_l+i\omega_k}\frac{\chi_j}{\lambda_j-i\omega_k}\,.\label{eq:35A}
\end{SMEquations}
Exploiting Eq.~(\ref{eq:33A}) we rewrite Eq.~(\ref{eq:35A}) as
\begin{SMEquations}
\lim_{t\to\infty}\dot{I}_2^{(k)}(t)=\frac{c_k^2}{2mm_k}\coth\left(\frac{\beta\omega_k}{2}\right)\frac{J(\omega_k)}{m}\tilde{\chi}(\omega_k)\tilde{\chi}(-\omega_k)\,.\label{eq:36A}
\end{SMEquations}
From the properties of $\tilde{\chi}_{\mathrm{r}}(\omega)$ and $\tilde{\gamma}(\omega)$ it is known that~\cite{Carrega22}
\begin{SMEquations}
\frac{J(\omega)}{m}\left|\tilde{\chi}(\omega)\right|^2=\tilde{\chi}''(\omega)\,.\label{eq:37A}
\end{SMEquations}
Substituting this result in Eq.~(\ref{eq:36A}) and inspecting Eq.~(\ref{eq:34A}) one can finally recognize that
\begin{SMEquations}
\lim_{t\to\infty}\left[\dot{I}_1^{(k)}(t)+\dot{I}_2^{(k)}(t)\right]=0\,.
\end{SMEquations}

\bibliographystyle{naturemag}
\bibliography{biblio}